# Time reversed optical waves by arbitrary vector spatiotemporal field generation

Mickael Mounaix[1,†], Nicolas K. Fontaine[2,†], David T. Neilson[2], Roland Ryf[2],

Haoshuo Chen[2], Juan Carlos Alvarado-Zacarias[2] and Joel Carpenter[1*]

**Abstract :** Lossless linear wave propagation is symmetric in time, a principle which can be used to create time reversed waves. Such waves are special "pre-scattered" spatiotemporal fields, which propagate through a complex medium as if observing a scattering process in reverse, entering the medium as a complicated spatiotemporal field and arriving after propagation as a desired target field, such as a spatiotemporal focus. Time reversed waves have previously been demonstrated for relatively low frequency phenomena such as acoustics, water waves and microwaves. Many attempts have been made to extend these techniques into optics. However, the much higher frequencies of optics make for very different requirements. A fully time reversed wave is a volumetric field with arbitrary amplitude, phase and polarisation at every point in space and time. The creation of such fields has not previously been possible in optics. We demonstrate time reversed optical waves with a device capable of independently controlling all of light's classical degrees of freedom simultaneously. Such a class of ultrafast wavefront shaper is capable of generating a sequence of arbitrary 2D spatial/polarisation wavefronts at a bandwidth limited rate of 4.4 THz. This ability to manipulate the full field of an optical beam could be used to control both linear and nonlinear optical phenomena.

Introduction : In a time reversal experiment[1–6] the spatiotemporal field to be recreated is often a short pulse originating from a small focused spot. After potentially undergoing a complicated scattering process the far-field of this source is recorded by an array of transducers/antennas. The time axis of these signals is then flipped and replayed through the array to regenerate the spatially and temporally focused source. This can be extended to a transfer matrix based approach[7–12] whereby an array of sources is characterised allowing arbitrary superpositions of those sources to be regenerated. Time reversal processing can be performed either by physically back-propagating signals through the medium, or by numerically back-propagating signals using the conjugate transpose of the transfer matrix measured in the forward direction. An advantage of a transfer matrix approach is the ability to deliver arbitrary spatiotemporal fields to the target, without first having to physically generate, back-propagate and measure the required input field, meaning experimental error associated with physical field generation is eliminated, and any field can be synthesised

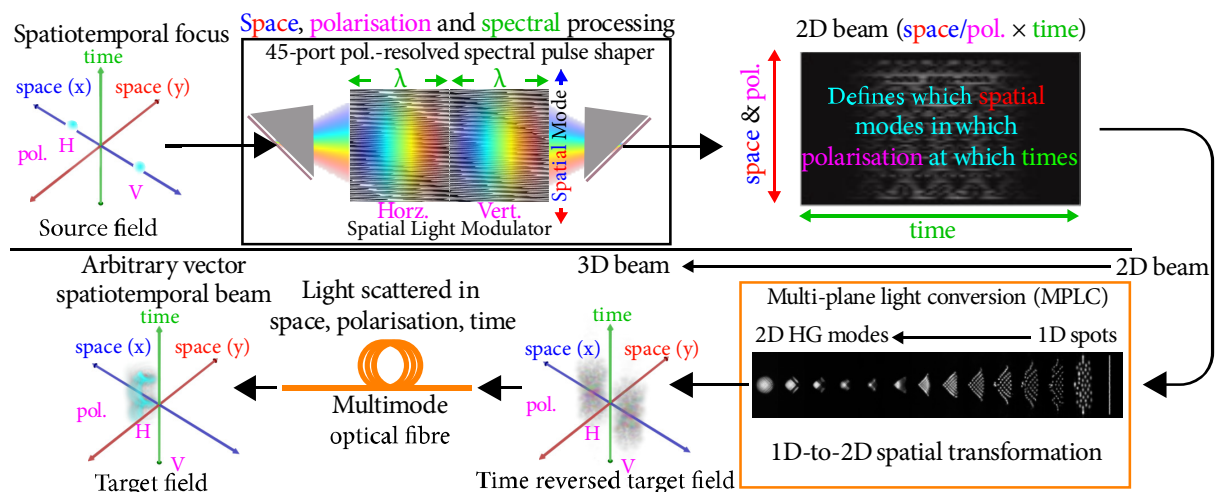

**Fig. 1 | Simplified schematic of a device capable of mapping an input vector spatiotemporal field onto an arbitrary vector spatiotemporal output field.** Amplitude, phase, spatial mode, polarisation and spectral/temporal degrees of freedom can all be independently addressed simultaneously through the programming of the spatial light modulator. 90 spatial/polarisation modes can be independently controlled over 4.4 THz at a resolution of approximately 15 GHz, making a total of approximately 26,000 spatiospectral modes. Source, time reversed and target fields have polarisation components illustrated spatially separated for clarity, but are co-located. Optical system animated in detail in second online video (14:00) in Supplementary Note 1.

[1] School of Information Technology and Electrical Engineering, The University of Queensland, Brisbane, QLD, 4072, Australia. [2]Nokia Bell Labs, 791 Holmdel Rd., Holmdel, NJ 07722, USA  †These authors contributed equally: Nicolas K. Fontaine, Mickael Mounaix. *e-mail: j.carpenter@uq.edu.au

from a small number of measurements, not just fields which have been previously measured.

Low frequency phenomena such as acoustics, water waves and microwaves, are within reach of electrical digitisers and signal generators that can record and generate the required fields directly in the time domain. When working with broadband sources at optical frequencies, such as femtosecond lasers, the electric field cannot be directly measured or manipulated in the time domain as it can for acoustics or microwaves. Hence extending time reversal techniques into optics requires a different approach.

For a given target field after propagation through a complex medium, the exact corresponding time reversed wave in the general case is a completely arbitrary function of space, time and polarisation. The scattering process itself need not create waves with any correlations between the spatial, temporal and polarisation degrees of freedom, and the therefore any device capable of generating these time reversed fields must be similarly unrestricted. The generation of such fields requires independent control of the impulse response of every spatial and polarisation mode supported by the medium. In this sense, it is the ultimate form of linear wave control as it requires all the classical linear properties of the wave to be controlled independently and simultaneously. Previous experiments in optics[10,13,14] have demonstrated spatial control[9,15–21], temporal control [22–25] or some limited combination of both[26–33]. These demonstrations have various implementations, however they all share an inability to simultaneously control 2D space, time/frequency and polarisation as completely independent degrees of freedom. In previous demonstrations, each property of light does not map to its own spatially separated position on the programmable wavefront control device (typically an SLM) where it can be independently the transverse spatial properties of a beam are two-dimensional. In Cartesian coordinates for example, the transverse structure of a beam is a function of both *x* and *y*. Hence there are three dimensions of required control per polarisation, which in the simplest case must be addressed by the two-dimensional surface of the SLM. This dimensional mismatch is an important reason why fully time reversed waves in optics have not previously been demonstrated. Previous work used the two dimensions of the SLM to control the two spatial dimensions[9], 1 spatial dimension and 1 spectral dimension[27,34], or some other partial combination of the spatial and temporal degrees of freedom [13,22,26,29,30,32,35].

In this work as summarised in Fig. 1, we employ a multi-plane light conversion (MPLC) device[38–40] in combination with a polarisation-resolved[41,42] multi-port[37] spectral pulse shaper[36] in order to control all three dimensions (2 space, 1 frequency/time) for both polarisations on a two-dimensional SLM. The MPLC device performs a spatial transformation which maps a one-dimensional array of 45 Gaussian spots to a two-dimensional set of 45 Hermite-Gaussian (HG) modes[39,40]. This enables two spatial dimensions of the output beam to be controlled using a single spatial dimension of the SLM, which leaves the other spatial dimension of the SLM for control of the spectral/temporal degree of freedom. The spectral pulse shaper steers light to a set of 45 discrete spot positions in a 1D array, which through the MPLC transformation, excite the corresponding 2D HG modes at the input of the multimode fibre. Addressing the output beam in the HG basis naturally supports applications in both free-space and fibre optics. Compared to spatial bases of 2D spot arrays, the HG basis also has the advantage of no regions of dead-space where light cannot be delivered in the near and far-fields.

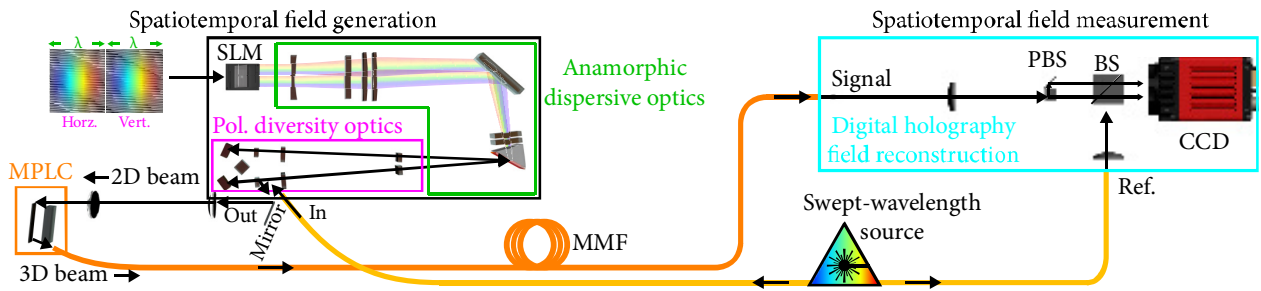

**Fig. 2 | Schematic of spatiotemporal field generation and characterisation apparatus.** A polarisation and spatially resolved spectral pulse shaper for generating arbitrary vector spatiotemporal states, in conjunction with a swept-wavelength digital holography system for characterisation. All characterisation and results are measured in the frequency domain.

controlled. For example, all spectral components may map to the same position on the SLM[26], or one spatial axis cannot be controlled independently of the spectral axis[27,34,35], and it is also common for polarisation to be neglected entirely. Due to the high frequencies and bandwidths, the manipulation of ultrashort optical pulses is typically performed in the frequency domain using spectral pulse shapers[36]. In the frequency domain, time reversal is broadband phase conjugation[13,27], and requires the amplitude and phase of every spatial mode to be independently controllable as a function of wavelength. A spectral pulse shaper based on a 2D spatial light modulator (SLM) can control the frequency response of multiple spots in a linear array simultaneously[37]. In such a system, one-axis of the SLM is assigned to the wavelength degrees of freedom and the other axis the one-dimensional spatial degrees of freedom, consisting of the output linear spot array. However,

## Results

**Experimental Setup.** A schematic of the device and the associated characterisation apparatus is summarised in Fig. 2. Detailed explanations and visualisations are available in the Supplementary Information and online videos listed in Supplementary Note 1. The source field enters the device through a single-mode fibre (SMF) in some arbitrary polarisation-dependent temporal state. Inside the device, this source field will be split amongst many optical paths by the SLM, creating different combinations of spatial modes and delays. This redistribution of light transforms the source into the desired spatiotemporal or spatiospectral state at the output. In these demonstrations, the source corresponds in the time-domain with a bandwidth-limited sinc pulse (4.4 THz of rectangular bandwidth centred at 1551.4 nm), linearly polarised at 45 degrees with respect to the SLM. This source field enters the polarisation-resolved multi-port spectral pulse shaper where it is mapped onto the surface of the SLM (Holoeye PLUTO II) through polarisation diversity optics and anamorphic dispersive optics. Through these optics, the horizontally and vertically polarised components of the beam are separated onto the left and right side of the SLM respectively. Within each beam for each polarisation component, the spectral components are dispersed across the $x$-axis (wide axis) of the SLM. Applying a phase tilt along this spectral axis ($x$-axis) will steer the beam back towards the output array along longer or shorter paths through the grating, creating controllable delay[36]. In Fig. 2, this corresponds with steering in the plane of the page. The device operates between 1533.94 nm and 1569.27 nm, corresponding to 4.4 THz of optical bandwidth. The width of each spectral component on the SLM is approximately 3 pixels or 15 GHz. Corresponding with a maximum of approximately 300 spectral modes of control for each of the 90 spatial/polarisation modes. Applying a phase tilt along the $y$-axis (short axis) of the SLM will steer the beam along the 1D array of Gaussian spots at the input to the MPLC[39,43]. In Fig. 2, this corresponds with steering in/out of the plane of the page. Each of the 45 spots in the 1D array is transformed to a particular HG mode through the 14 phase planes of the MPLC device for both polarisation components. In this way, both Cartesian indices ($m,n$) of the HG basis set can be addressed by steering light along the 1D array using the SLM. By programming more complicated phase masks onto the SLM it is possible to redistribute light amongst these two axes corresponding to the spatial/polarisation and spectral/temporal degrees of freedom, allowing arbitrary spatial/polarisation modes to be assigned to arbitrary frequencies or delays. The device is attached to a 5 m length of graded-index 50 μm core diameter multimode fibre (MMF)[44] which will be used as a complex medium through which a desired vector spatiotemporal field is to be generated. The fibre supports the same number of modes as the spatiotemporal beam shaper (90 spatial/polarisation modes) with a delay spread of approximately 0.15 ps/m[44]. Due to modal dispersion, a spatiotemporal state input to the fibre arrives at the other end in a different spatiotemporal state. As the entire delay spread of all spatial and polarisation modes this fibre supports are addressable, no path exists through the fibre which cannot be time reversed.

A complete linear description of the device and the attached MMF is acquired using swept-wavelength digital holography[39]. This linear description consists of a set of 90×90 frequency-dependent complex matrices, which maps each of the 90 spatial/polarisation input modes to each of the 90 output modes, in both amplitude and phase as a function of optical frequency. Through these matrices, any spatiotemporal input can be mapped to any spatiotemporal output in both directions. To measure these matrices, each spatial mode in each polarisation is selectively excited at the input of the MMF one-at-a-time by the SLM. Then, as the wavelength of the source is swept, the digital holography system measures the optical field at the output of the fibre, and extracts the amplitude and phase for each output mode as a function of frequency. The desired spatiotemporal state to be generated at the output of the fibre is specified as a wavelength-dependent complex vector, which when back-propagated through the measured matrices yields the corresponding input spatiotemporal state. A phase mask is then calculated to generate this state using a modified Gerchberg-Saxton algorithm and displayed on the SLM. Swept-wavelength digital holography is then performed to characterise the resulting spatiotemporal output state.

In traditional holography, a two-dimensional diffractive element encodes the complex amplitude of a two-dimensional wavefront, which can be recreated by illuminating the element with a spatial reference beam. This new device can be thought of as an extension of this to an extra dimension; a three-dimensional diffractive element which when illuminated with a reference pulse in a reference spatial mode, will reconstruct a fully volumetric optical field (2 transverse space and 1 time/longitudinal space). It is a type of reprogrammable space-time hologram[45].

This device is capable of generating types of optical beams, with full control of the spatial, spectral/temporal and polarisation properties and the implementation of various mappings between these properties, for example, arbitrarily polarised focused spots or singularities[46,47] tracing arbitrary trajectories through space and time or frequency[48], or indeed any arbitrary vector spatial fields generated at arbitrary wavelengths or delays. In this way the device can also be thought of as a kind of ultrafast wavefront shaper, capable of generating a sequence of approximately 90 independent wavefronts at a rate of 4.4 THz. For a traditional continuous wave (CW) beam, the wavefront in one plane specifies the wavefront at all other points in space ahead and behind the wavefront. However these new beam types have an additional dimension of control. For a fixed point in time, or alternatively a fixed plane in space along the optic axis, like for example the focal plane, the wavefronts ahead and behind can all be independently controlled and unrelated.

Spatial wavefront manipulation and spectral pulse shaping already have broad applicability and the ability to perform both simultaneously by this device could have many applications within linear and nonlinear optics, for example, the control of light propagation through complex media for applications such as imaging, which is analogous to previous demonstrations for lower frequency phenomena. However, optical waves are quite different from acoustics, microwaves and water waves, not only in terms of wavelength, frequency and bandwidth, but also particularly with respect to interaction with matter. Hence, this new type of control in optics could open up many possibilities that are not just generalisations of previous demonstrations for lower frequency phenomena, with applications such as nonlinear microscopy[49], micromachining[50], quantum optics[51], optical trapping[52], nanophotonics and plasmonics[53], optical amplification[54] and other new nonlinear spatiotemporal phenomena, interactions and sources[55-57].

**Experimental Results.** Fig. 3 contains various examples of spatiospectral and spatiotemporal control, with additional examples available in the Supplementary Information. The measured 3D optical fields (2D space and 1D time/frequency) are plotted as a sequence of 2D fields, as well as volumetric renderings. The plots contain both amplitude and phase information for these fields for both polarisation components. Examples, such as the spatiotemporal foci of Fig. 3b and Fig. 3d, are similar to typical first demonstrations of time reversal in other wave phenomena, which is a use case of practical application in imaging and nonlinear optics. However most examples were chosen to be illustrative of the system's ability to generate beams with arbitrary control over space, time/frequency and polarisation, rather than any specific use case. All beam types are characterised in the frequency domain.

For spatiotemporal beam demonstrations, the presented results are Fourier transformed into the time domain from the measured frequency dependent fields measured using swept-wavelength digital holography.

In Fig. 3a the device of Fig. 2 is used as a programmable dispersive element, which can perform arbitrary mappings between wavelength and 2D space/polarisation. The horizontal and vertical polarisations cycle through spatial fields corresponding with letters of the Latin alphabet in forward and reverse alphabetical order respectively. From these measured optical fields of Fig. 3a it is possible to see that the spatial amplitude and phase in both polarisations can be controlled as a function of wavelength. The overall spectral phase is near constant as a function of wavelength and hence all letters arrive at the same delay.

The remaining examples of Fig. 3 all demonstrate spatiotemporal beams. In these examples, the principle of operation is much the same as the spatiospectral control example of Fig. 3a, except the desired spatial/polarisation output states are specified in the time domain. For temporal control, it is critical that not only are the correct superpositions of spatial/polarisation modes excited at each wavelength, but also the relative phase between the wavelengths must be correct in order to generate the desired temporal features. Fig. 3b is a relatively simple spatiotemporal demonstration; a polarised diffraction limited focus of 230 fs temporal duration (corresponding to the bandwidth of the device, 4.4 THz). As discussed in more detail in the Supplementary Information (Supplementary Figure 16), at 0 ps delay, the spatial focus is 86% of the peak theoretical intensity achievable using 45 HG modes, and 84% of the theoretically achievable power for a rectangular spectrum of 4.4 THz is delivered to the 0 ps delay. Fig. 3c is a similar demonstration of a more complicated spatial field, a vertically-polarised "smiley face", which has been generated at a desired delay of 5 ps. Fig. 3d in an example of two orthogonally polarised focal spots that are independently tracing out spirals in time across a delay spread of 20 ps. This can be seen from the sequence of 11 measured fields shown in Fig. 3d and from the volumetric rendering of the field to the right of Fig. 3d. This illustrates how the device can be used to trace focused spots along arbitrary trajectories at a rate limited by the optical bandwidth (4.4 THz). This could be used for raster scanning or other kinds of sampling at ultrafast rates. Fig. 3e is another spatiotemporal demonstration, similar to Fig. 3d, but which maps more complicated spatial fields representing numerals to different delays across an 8 ps delay spread. Fig 3f is an example which illustrates a spatial and polarisation state change as a function of delay. A radially polarised "clock hand". That is, a spatial field consisting of a linearly polarised line pointing outwards from the centre of the fibre core, which rotates as a function of delay. This kind of control of space, time and polarisation independently could be of use to high-power applications working at high numerical apertures, such as micromachining[50] or optical trapping[52]. The examples of Fig. 3g and 3h are designed to represent more recognisable 3D objects. Fig. 3g, the "arrow of time", corresponds with a light field literally shaped like a 3D arrow. Launching a pulse into the input of the spectral pulse shaper of Fig. 2 would result in a 3D arrow shaped beam, pointed backwards along the time/propagation axis, flying into the CCD camera of the characterisation apparatus. Similarly, Fig. 3f is a simple 3D light rendering of an Eiffel tower shaped beam flying base-first along the propagation axis. Further experimental results are available in the Supplementary Information, particularly in the online videos, including additional characterisations such as spatially and temporally dependent losses, as well as tests of performance versus number of addressed spatiospectral modes.

The imperfections in the device can be organised into three types. First, there is frequency-dependent loss (Supplementary Figure 8). The frequency response of the device is not perfectly flat over the full 4.4 THz range, which makes for a slight broadening of the temporal features. Integrating under the average frequency response curve over all 90 spatial/polarisation modes, yields a bandwidth that is 83% compared with a flat 4.4 THz response. This is consistent with the 84% value for the specific example of the spatiotemporal focus presented in Fig. 3b and Supplementary Figure S16. The second type of imperfection is mode dependent loss (Supplementary Figure 8). Each of the 90 spatial/polarisation modes do not have the same loss, which leads to a loss of spatial/polarisation detail. This effect averaged over all frequencies and all possible spatial/polarisation target fields yields a spatial/polarisation fidelity of 83%. That is, when maximising the power delivered to the target state, and assuming no additional imperfections that were not present when the transfer matrices were measured, on average 83% of the total output power would be in the target spatial/polarisation state. A thorough investigation of the distribution of spatial fidelities over frequency and target state is provided in Supplementary Figure 17. Both frequency and spatial/polarisation dependent losses could at least to some extent be calibrated out by trading off total power delivered to the target state, in favour of a higher proportion of the output power delivered to the target state. In practice, that would mean redistributing and/or attenuating the power of spatial and spectral degrees of freedom to flatten the overall response.

The third imperfection type relates to the accuracy at which arbitrary spatiotemporal /spatiospectral states can be generated. This is largely about the ability of the SLM to accurately represent the required phase masks, in the presence of limitations such as pixel crosstalk and finite spatial resolution. This is discussed in more detail in the Supplementary Information (Supplementary Figure 11 to 15). For the most complicated spatiospectral states that require all 90 spatial and polarisation modes to be excited equally across the entire addressable frequency band, the proportion of the total output power delivered to the target state is typically 86% where the spectral features are wider than 30 GHz ($90 \times (4400/30) = 13200$ spatiospectral modes). This value drops to 77% for 20 GHz features (19800 spatiospectral modes) and 39% for 10 GHz features (39600 spatiospectral modes), which exceeds the spectral width of the beams on the SLM (~15 GHz).

We have demonstrated a system capable of measuring and generating arbitrary vector spatiospectral optical fields. This device is able to simultaneously control all classical linear degrees of freedom in an optical beam independently, enabling full time reversed optical waves to be generated through complex media as well as spatiotemporal control of light more generally for applications such as imaging, nonlinear optics and micromanipulation. Alternate designs and options for scaling this device to higher number of spatial modes, longer delays, finer temporal features, and/or lower loss, are discussed in the Supplementary Information.

**Data availability** Data available on request from the authors.

**Acknowledgements** We acknowledge the Discovery (DP170101400, DE180100009) program of the Australian Research Council, and the support of NVIDIA Corporation with the donation of the GPUs used for this research. We acknowledge Martin Plöschner for helpful discussions and Marcos Maestre Morote for work on the swept-wavelength trigger circuitry.

**Author Contributions** Experiments performed by M.M and N.F with assistance from J.C.A-V, R.R and H.C. Optical system designed by N.F, D.N, M.M and J.C. Spectral pulse shaper section designed by D.N and N.F, MPLC by N.F and J.C, spatiotemporal holograms designed by M.M., N.F. and J.C. Optical system assembled by N.F. Data analysis by M.M, N.F and J.C. Manuscript written by J.C, M.M and N.F with input from all authors. J.C conceived and supervised the project.

**Competing interests** The authors declare no competing interests.

**Additional information**

**Correspondence and requests for materials** should be addressed to J.C. (j.carpenter@uq.edu.au).

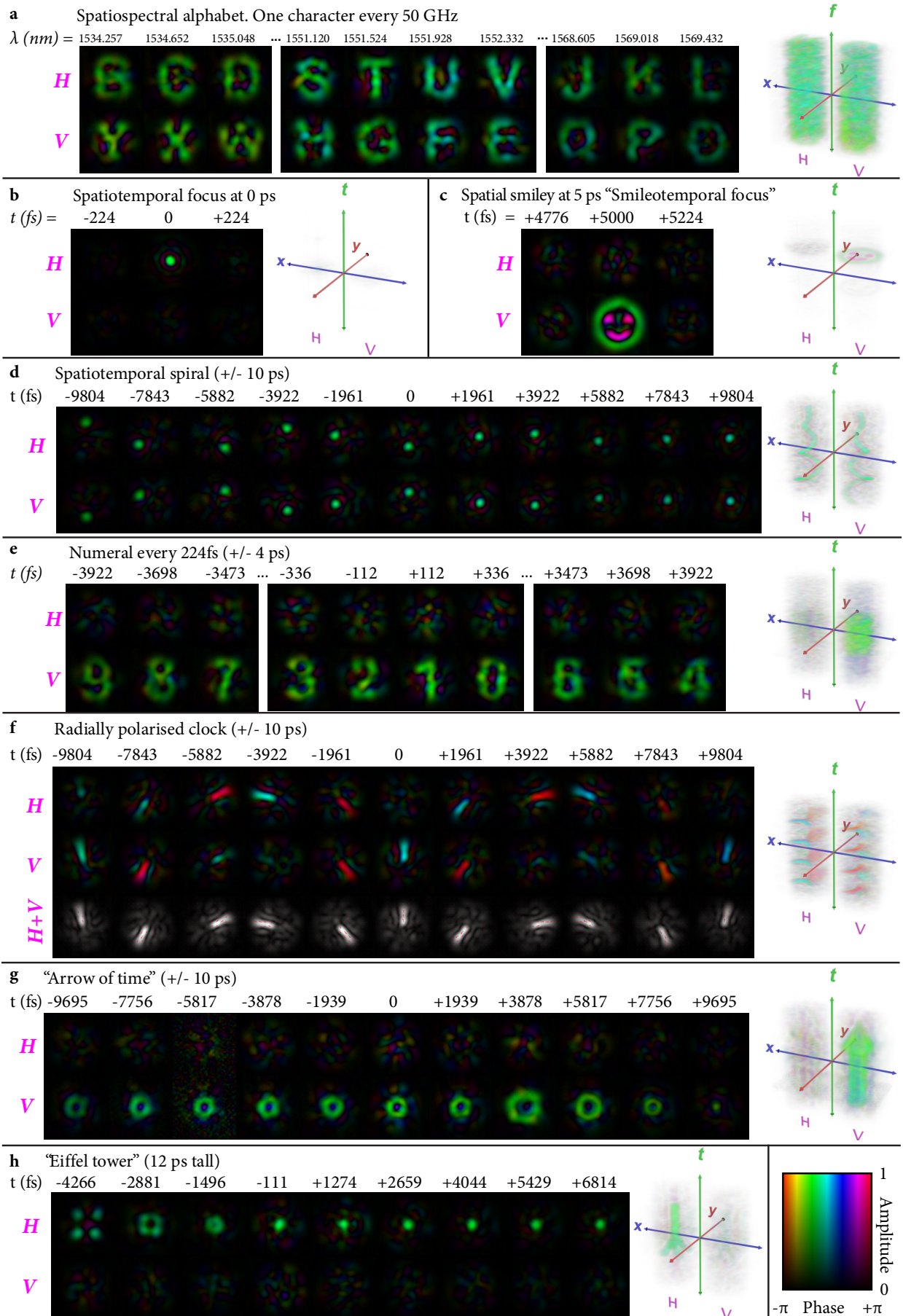

**Fig 3 | Selection of vector spatiospectral and spatiotemporal states measured at the distal end of the multimode optical fibre.** Various examples illustrating control of the spatial amplitude, phase and polarisation of a beam as a function of frequency or time. The examples are shown as a sample sequence of measured optical fields, as well as a volumetric rendering of the field as function of space, time/frequency and polarisation. (a) Spatiospectral demonstration. (b-h) Spatiotemporal demonstrations are Fourier transforms of the measured optical fields using swept-wavelength digital holography. Further examples and animations are available in the Supplementary Information.

# Time reversed optical waves by arbitrary vector spatiotemporal field generation

Mickael Mounaix[1], Nicolas K. Fontaine[2], David T. Neilson[2], Roland Ryf[2], Haoshuo Chen[2],

Juan Carlos Alvarado-Zacarias[2] and Joel Carpenter[1]

[1] School of Information Technology and Electrical Engineering, The University of Queensland, Brisbane, QLD, 4072, Australia.

[2] Nokia Bell Labs, 791 Holmdel Rd., Holmdel, NJ 07722, USA

## Supplementary Note 1 : Videos

Two videos are included which complement the content of the document itself. The first is a 5 minute summary of this work aimed at a more general audience, and the second, a 75 minute detailed technical summary of all work presented in the main document and Supplementary Information.

Video 1, *General Audience Summary (5 mins)* : https://youtu.be/1WcIejZd__w

Video 2, *Detailed Technical Summary (75 mins)* : https://youtu.be/9hVEJvfWjRQ

***Technical Video Table of Contents {minutes:seconds}***

*Basic principles and examples of time reversal*
- 00:32     Time reversal in water waves
- 02:39     Time reversal in acoustics
- 03:13     Time reversal in microwaves
- 03:53     Focusing through scattering objects

*Previous work on extending time reversal techniques into optics*
- 05:08     Frequencies and bandwidth of optics
- 06:01     Spatial-only control
- 07:47     Space-time coupling in scattering media
- 09:36     Existing types of wave control in optics (spatial, temporal, spectral etc)
- 11:52     Time-domain spatiotemporal field generation (acoustics/microwaves and the difficulty of optics)
- 13:06     Frequency-domain spatiotemporal field generation

*Device details and applications*
- 14:00     Device principles of operation
- 18:51     1D-to-2D transformation by MPLC
- 19:40     Device as ultrafast wavefront generator
- 21:02     Advantage of Hermite-Gaussian spatial basis over 2D spot array
- 21:21     Device as volumetric hologram
- 21:54     Arbitrary spatiotemporal field generation and its applications
- 24:16     Detailed schematic walk-through

*Characterisation apparatus*
- 25:41     Measurement of the transfer matrices, summary
- 25:54     Fibre-under-test
- 26:25     Measurement of the transfer matrices, detailed
- 28:48     Backpropagation
- 29:31     Spatiotemporal characterisation of states

*Phase mask calculation*

# Supplementary Methods

## Experimental apparatus

*Summary*

*(video timestamp : 24m14s)*

The experimental apparatus can be thought of in two parts; the spatially and polarisation resolved spectral pulse shaper itself, and the swept-wavelength digital holography system used to characterise it. Both are shown in Fig. 2 of the main document which for convenience is repeated here as Supplementary Figure 1.

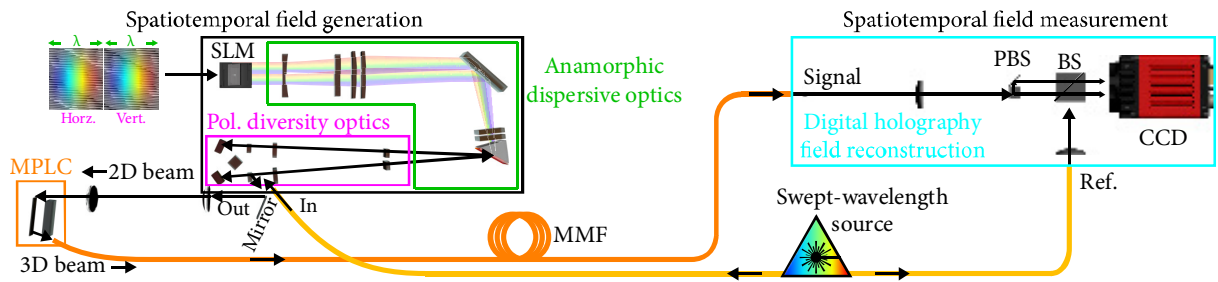

**Supplementary Figure 1 | Schematic of optical time reversal system and characterisation apparatus.** A polarisation and spatially resolved spectral pulse shaper for generating arbitrary vector spatiotemporal states, in conjunction with a swept-wavelength digital holography system for characterisation.

The spatially and polarisation resolved spectral pulse shaper, is illustrated on the left-side of Supplementary Figure 1 and consists of two optical sub-systems. First, the polarisation-resolved multi-port spectral pulse shaper sub-system. This consists of polarisation diversity optics (pink in Supplementary Figure 1), dispersive and beam resizing optics (green), and the spatial light modulator (SLM). The operation principles of this portion of the device are similar to that of other spectral pulse shapers[1] or wavelength-selective switches[2], albeit with the unusual inclusion of polarisation-diverse processing and a relatively large number of addressable single-mode Gaussian output ports. The output ports in this case are a 1D array of 45 Gaussian spots, with each spot being addressable by steering light in the direction in/out of the page as of Supplementary Figure 1 by applying the appropriate phase ramp across the vertical direction of the SLM. Light from the spectral pulse shaper section is imaged through a 4$f$ telescope ($f$=10cm) onto the multi-plane light conversion (MPLC) device[3]. This MPLC sub-system maps the 1D array of Gaussian spots onto a 2D set of HG modes through 14 planes of phase manipulation. For a given 2D phase mask programmed onto the SLM, a corresponding 2D field is generated in the Fourier plane, approximately at the grism. This 2D field in turn becomes a 3D field through the 1D-to-2D transformation of the MPLC device. This field is coupled into the multimode optical fibre (MMF)[4] at the output of the MPLC. As light propagates through the MMF it becomes redistributed spatially and temporally due to mode coupling and mode dispersion along the length. At the output of the MMF this field propagates into the swept-wavelength digital holography portion of the setup. This portion is used to characterise the spatial, polarisation and spectral/temporal properties of the beams[5,6]. Both to measure the optical transfer function[7] of the device and the fibre under test, as well as to characterise the spatiotemporal and spatiospectral beams being generated out the output of the fibre.

The swept-wavelength laser used is a NewSpec TLM-8700-L-CL. The CCD camera is an Allied Vision Goldeye CL-008. An external Mach-Zehnder interferometer samples light from the TLM-8700 as the laser is swept and through additional circuitry generates camera triggers at fixed frequency spacing. Which for the results presented here is 10.0675 GHz spaced frequency samples from 195.510 THz to 191.000 THz. The swept-wavelength digital holography characterisation procedure is similar to that explained in previous work[5,6].

*Spectral pulse shaper sub-system*

The spectral pulse shaper sub-system design schematic is illustrated in Supplementary Figure 2, in addition to a photograph of the sub-system in Supplementary Figure 3. The source light enters from the TLM-8700 swept-wavelength laser source as a single-mode optical fibre (SMF-28) attached to a microlens such that the source beam enters the optical system with a mode-field diameter (MFD) of 60 μm. The same beam diameter as the spots in the 1D array of the MPLC device which are optically in the same plane as the source. The input light has been aligned by a fibre polarisation controller attached to the TLM-8700 such that approximately equal power is in the horizontal and vertical polarisation components with respect to the optics of the pulse shaper at the output of the MMF. Hence input power will exit both ports of the polarising beam splitter (PBS) and propagate through spatially separated polarisation diversity optics for each polarisation component. M1h and M1v are simply dielectric mirrors. L1h/v and L3h/v are a set of identical cylindrical lenses which image the input source spot onto the grism in the *y*-direction (in/out of page, spatial axis of the device). L2h/v are cylindrical lenses along the orthogonal direction, collimating the input source spot onto the grism in the *x*-direction (plane of the page, spectral/temporal axis of the device). That is, the spot at the grating is small in the spatial direction, and wide in the temporal direction. When Fourier transformed onto the SLM for each spectral component, this becomes a large beam in the spatial direction and narrow in the spectral direction, corresponding with the ability to address a large number of spatial ports and large delays. HWP1 is a half-wave plate to align one of the incoming polarisation components (H) to that of the grating. For the other polarisation component (V), which is already aligned optimally for the grating, L0 is simply a flat block of glass to maintain optical path length matching between the polarisation components. Light is then incident on the grism and starts to disperse along the spectral/temporal axis (plane of the page). Together L4, L5, L6, L7 and L8 perform some resizing of the beams in the *x* and *y* direction, but ultimately are setup to place the SLM in the Fourier plane of the grism. M2 and M3 are simply dielectric mirrors and HWP2 aligns the incoming polarisation from the grism to that required by the SLM.

Programming a phase mask onto the SLM will cause light to be redistributed amongst the spatial and spectral/temporal axes of the device. Steering light to different positions along the spatial axis (*y*-axis, in/out of the page) of the grating correspond with different Hermite-Gaussian (HG) modes at the output of the MPLC. Different positions along the temporal axis (*x*-axis, plane of the page) correspond with longer or shorter path lengths traced back through the outer or inner edges of the grism. Light retraces its path back through HWP1/L0, L3, L2, L1, M1, and the PBS towards the input source. However due to a slight deliberate tilt, light through the return path is not coupled back into the laser source, but instead, is picked off with a mirror (M4) and routed towards the MPLC sub-system.

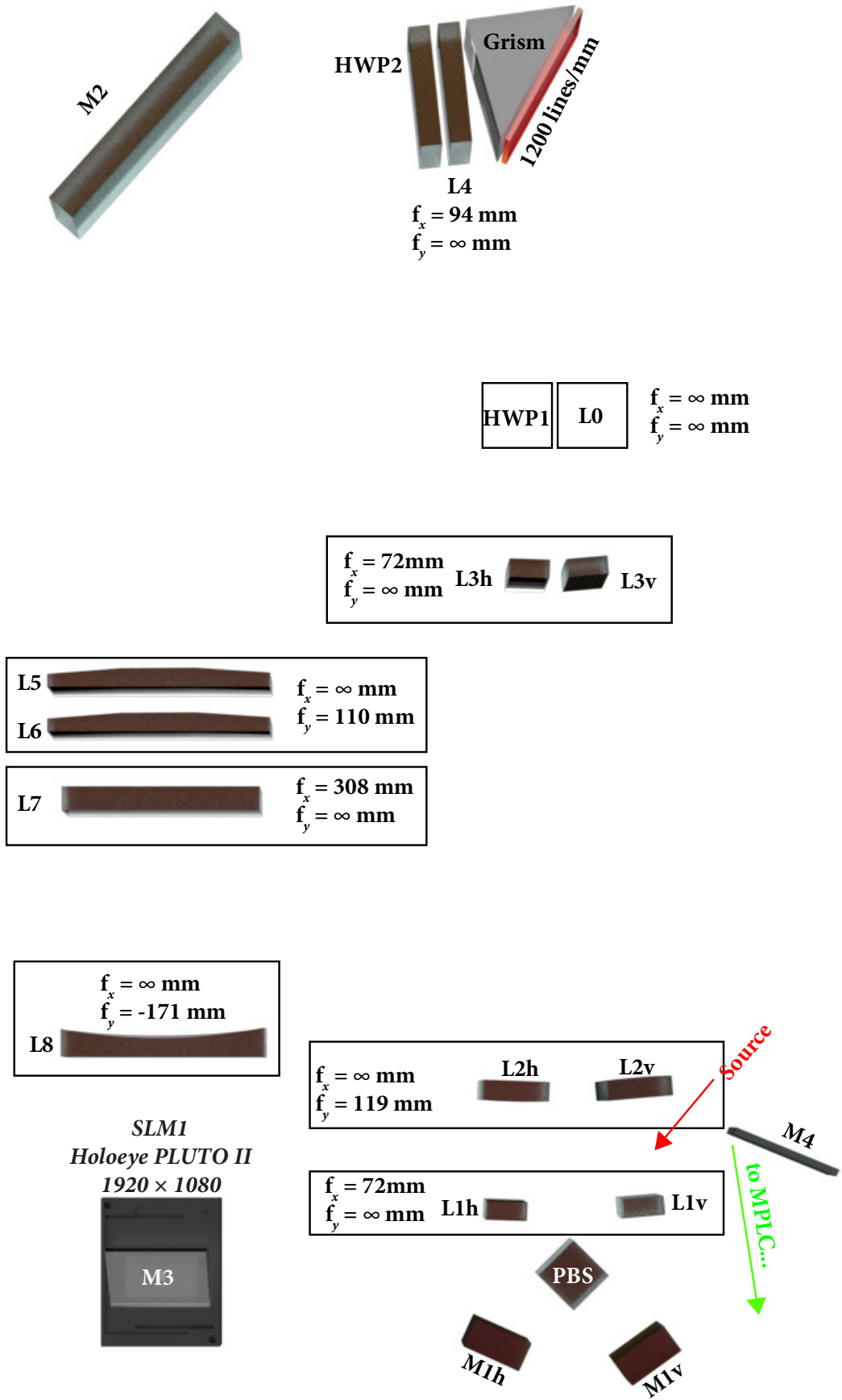

**Supplementary Figure 2 | Design schematic of the multi-port polarisation-resolved spectral pulse shaper sub-system**

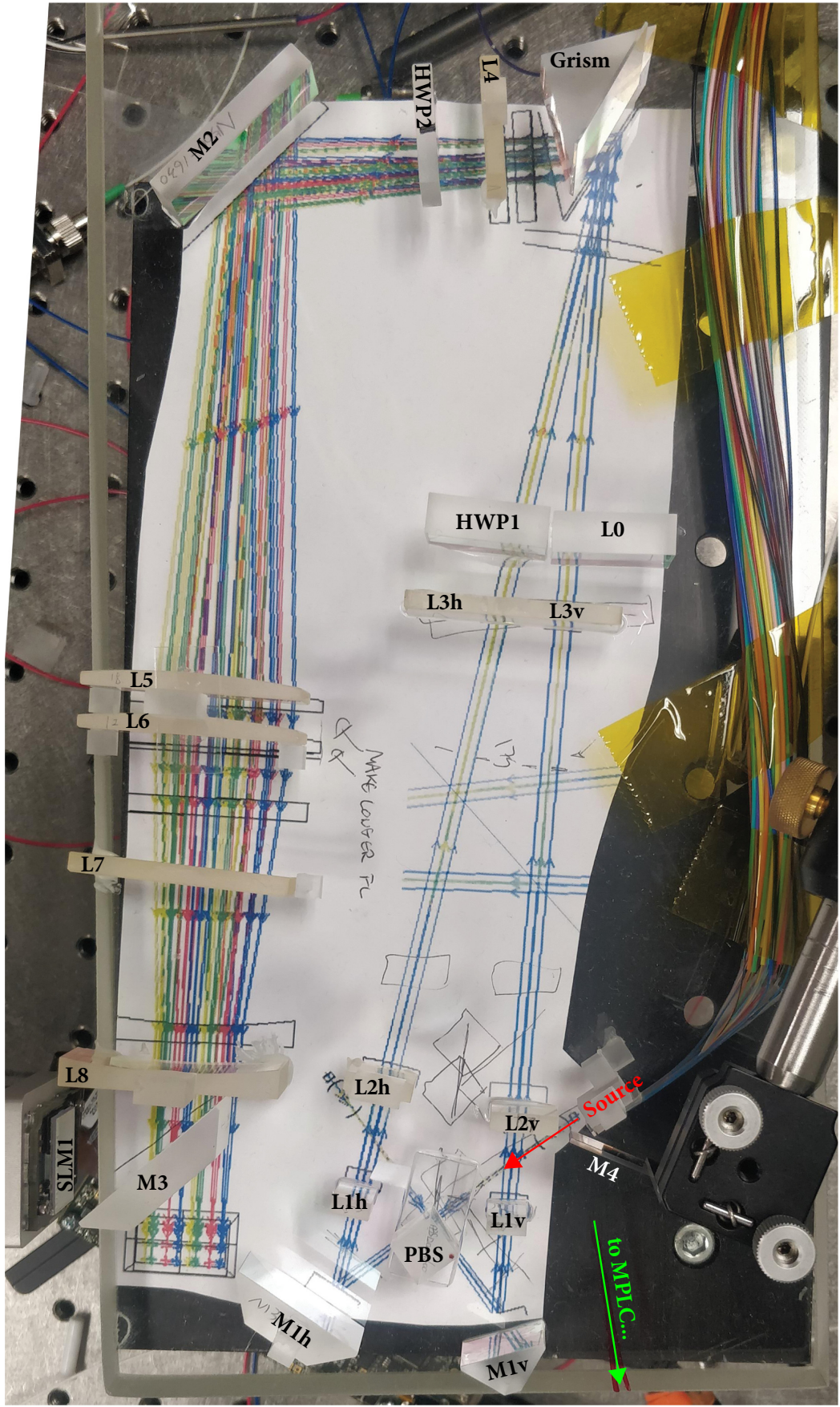

**Supplementary Figure 3 | Photograph of the multi-port polarisation-resolved spectral pulse shaper sub-system**

*Multi-plane light conversion sub-system*

As shown in Supplementary Figure 4, light is routed from the spectral pulse shaper sub-system to the MPLC sub-system by way of a 4*f* telescope (*f*=10cm). The MPLC device itself[3,8] consists of 14 lithographically etched phase planes which map a 1D array of 45 Gaussian spots (60 μm MFD, 127 μm pitch), to 45 Hermite-Gaussian modes of MFD 592 μm. Which are in turn coupled to the 50 μm diameter core graded index multimode fibre[4] through a *f*=4.51 mm lens. Further details on the MPLC design available here[3], as well as
https://www.youtube.com/watch?v=iXjFQj15Xzg&t=21m47s

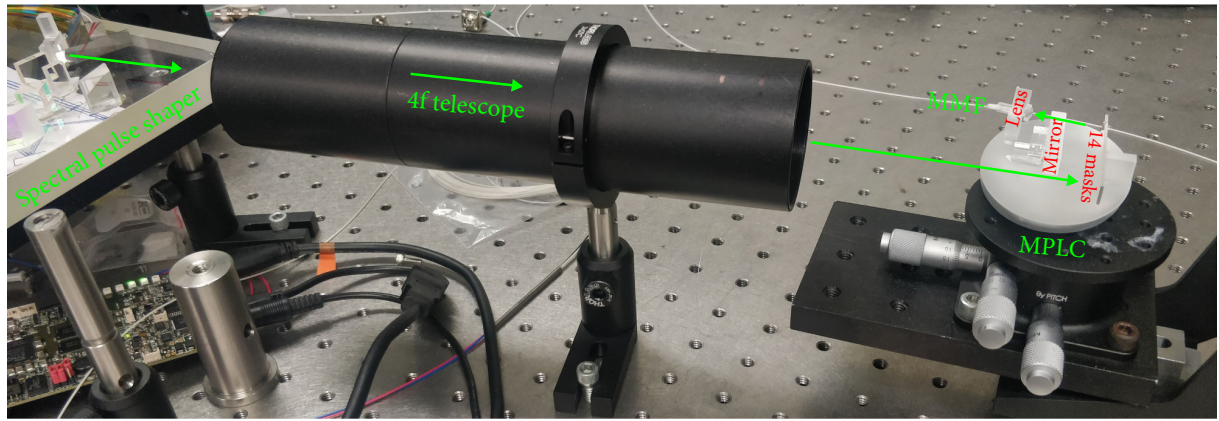

**Supplementary Figure 4 | Photograph of the lens relay imaging light from the spectral pulse shaper sub-system to the MPLC sub-system**

*Digital holography sub-system*

The digital holography sub-system is the characterisation apparatus used to measure the optical fields exiting the multimode fibre at the distal end. In off-axis digital holography, the intensity of the interference between the beam being characterised (image of the MMF core facet) and a reference beam (tilted quasi-plane wave) are measured. From this interference pattern the full complex optical field can be recovered[9]. A similar setup has been used previously[5] and is explained in more detail here; https://www.youtube.com/watch?v=TqAr4UFtoqw&t=38m33s

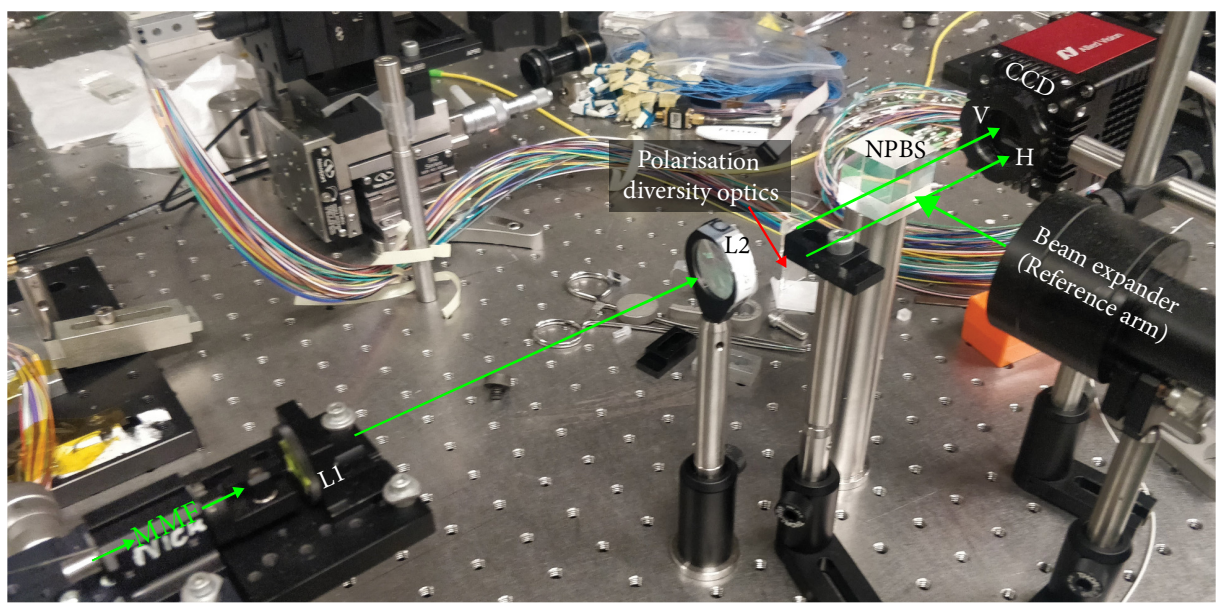

**Supplementary Figure 5 | Photograph of off-axis digital holography beam characterisation sub-system.**

The beam exiting the multimode fibre is magnified by a factor of 50 through L1 (*f*=3 mm) and L2 (*f*= 150 mm). The two polarisation components of the beam are displaced from one another using polarisation diversity optics before travelling through a non-polarising beamsplitter (NPBS) where these beams are interfered with the reference wave. The reference wave is a single-mode fibre attached to a beam expander coming from the same TLM-8700 laser source which enters the spectral pulse shaper. The interference between the reference wave and the horizontal and vertical beam components are recorded on the camera and then processed to recover the fields. The CCD camera is triggered by a *k*-clock. That is, an external Mach-Zehnder interferometer connected to additional circuitry which signals the camera to record images every 10.0675 GHz as the wavelength of the laser is swept. The measured optical fields in the basis of camera pixels are then overlapped with the first 55 Hermite-Gaussian modes to recover the complex amplitude of each spatial mode component in each polarisation. Ultimately the fibre and spectral pulse shaper only support 45 spatial modes per polarisation, however additional modes are decomposed for the purposes of alignment and checking the validity of results. The resulting data is then processed to remove the polarisation mode dispersion of the digital holography setup itself as well as removing residual group delay (6.05 ps) and chromatic dispersion (0.0127 ps$^2$) of the reference beam.

*Measurement of the optical transfer function (25m44s)*

The optical transfer function is a complete linear description of the system as a whole. In this case, it consists of a set of frequency-dependent complex transfer matrices which map any spatiotemporal input to any spatiotemporal output. It consists of 449 matrices (10.0675 GHz spaced samples across 4.5 THz). Each matrix maps the linear relationship between 90 spatial/polarisation modes as excited by the device itself (spectral pulse shaper in combination with MPLC and attached MMF) and the 110 spatial/polarisation modes measured on the digital holography system.

With regards to programming the SLM, each input mode is ultimately specified by a single number; the position of the corresponding spot in the 1D Gaussian array of the MPLC as seen from the plane of the SLM. As the SLM is in the Fourier plane with respect to the 1D array, each output spot corresponds with a particular angle of incidence on the SLM and the phase of the wavefront corresponding with this angle is what must be programmed onto the SLM to steer light to that position. Remembering that wavelengths are dispersed across the horizontal axis of the SLM, so the phase front for each spectral component will be slightly different, although the physical location (angle or position in the output array) is common to all wavelengths. Additional corrections such as aberrations and misalignments are also applied for best performance. For example, defocus, and the mapping of wavelength to pixel position on the SLM. This includes slight misalignments of the SLM axis with respect to the grating axis, as well as chromatic aberrations associated with the grism. A consistent basis for both the measurement of the transfer matrices, as well as the calculation of the required phase masks are required to successfully generate the desired output spatiotemporal and spatiospectral states of interest. The system is calibrated as a whole, by a single set of matrices which encapsulates the spectral pulse shaper, MPLC, MMF and digital holography sub-systems. As mentioned above, the polarisation mode dispersion, delay and chromatic dispersion of the digital holography apparatus is removed from the measurements.

Supplementary Figure 6 contains measured examples of the time reversal operator[10]. That is, the transfer matrix (*T*) of the system multiplied by its conjugate transpose (*T*$^*$). Supplementary Figure 6a contains the worst performing frequency (195.409 THz) and Supplementary Figure 6b corresponds with the best performing frequency (193.819 THz). Performance in this case is measured in terms of mode-dependent loss, which is the ratio of the maximum and minimum singular value, also known as the condition number. For an ideal device, the transfer matrix, *T*, would be unitary and the matrix, *TT*$^*$, would be the identity matrix, ignoring some overall loss which is common to all modes supported by the system. That is, an ideal device would have a transfer matrix where all the singular values are all the same, the inverse of *T* is equal to *T*$^*$ and the matrices of Supplementary Figure 6 would be diagonal, with all diagonal elements having the same value. Any deviation from this scenario means at least some degradation in the ability of the system to generate some fields that contain superpositions of spatial/polarisation components which the system cannot generate efficiently, if at all.

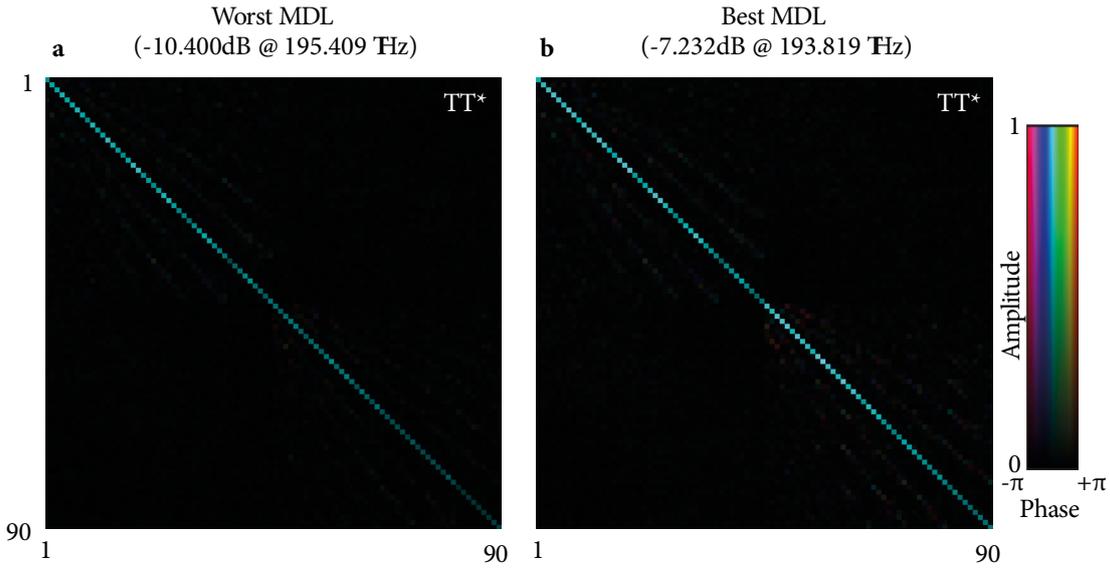

**Supplementary Figure 6 | Measured time reversal operators for the system. a,** Worst MDL frequency and **b,** Best MDL frequency.

*Calculation of the SLM hologram (31m18s)*

Generation of the desired spatiotemporal or spatiospectral field requires calculating a phase mask to display on the SLM which excites the correct amplitude and phase for each spatial and polarisation mode as a function of wavelength. How this 2D phase mask distributes light in 2D in the Fourier plane, approximately at the grating, defines how light from the source is ultimately distributed in space and time. The task of calculating the phase mask can be approached either in the spectral domain, or the temporal domain. We calculate the masks using a modified Gerchberg-Saxton (GS) algorithm[11,12]. The traditional GS algorithm is structured as a phase retrieval problem. It consists of two planes separated by a Fourier transform. In each plane, the intensity is a fixed goal and it is the phase in each plane that are the free parameters which the algorithm will attempt to optimise. In our application, the free parameters are arranged differently. The phase in the plane of the SLM is a free parameter, however the phase in the plane of the 1D array is not. All spots in the array must be excited with a specific amplitude and phase. Hence the free parameter in the plane of the 1D array is the region of space outside the array. Any light scattered to this region, will not be coupled into the MMF and hence not affect the beam quality, at the unavoidable expense of loss. For a single plane of phase manipulation it is impossible to guarantee lossless conversion. This is a property of all single plane holograms, not specific to this application. The reason for this can be understood by imagining the desired output spatiotemporal state back propagating through the system from the camera plane to the SLM plane. If a particular spatiotemporal state does not have an amplitude on the SLM that matches the amplitude of the Gaussian source beam coming from the laser in the opposite direction, there is nothing that a single phase mask can do to match the two beams (62m29s). At least two masks would be required, which can be imagined as one mask to redistribute the amplitude onto the next mask such that this second plane can match the phase (73m12s).

The modified GS algorithm iterates through the loop as follows;

a) Set the amplitude in the plane of the SLM to the amplitude of the Gaussian beam from the laser source. The phase in this plane remains unchanged from the previous iteration.
b) Fourier transform the plane of the SLM into the plane of the output 1D array.
c) Check the normalised overlap (beam quality) of the generated field with the goal field. Check the unnormalised overlap (loss). We want to know both how accurately the mask is exciting the modes with the correct amplitude and phase, but also how much power we are discarding to achieve this.

d) The currently generated field in the region of interest nearer the 1D array is then replaced with the ideal target field with some weighting. The relative weighting between the ideal target being applied inside the region of interest, and the background scattered light which is allowed to remain, defines the total loss of the hologram. If the background scatter is weighted too high, the overall loss will be high. If the background scatter is weighted too low, the algorithm does little on every iteration converging to a low quality result.
   e) The algorithm is continually iterated until a desired beam quality is achieved. Typically 95%.

For the experimental results shown here, we have calculated the masks as a set of $N$ 1D holograms corresponding to $N$ spectral components. In order to promote smooth 2D masks, the result from the 1D hologram for one wavelength is used as the starting point for the calculation of the next 1D hologram. An alternative approach of similar effect is to calculate the holograms in 2D using a target distribution specified in the time domain, with a spatial filter in the iteration loop to promote low spatial resolution solutions.

An example illustration is shown in Supplementary Figure 7. In this example, two focused spots are being created at the output of the fibre. One spot is horizontally polarised at 0 ps and the other is a vertically polarised spot at a different spatial location at a delay of -26 ps. Supplementary Figure 6a is the literal greyscale bitmap programmed onto the SLM which includes all corrections and calibrations. Supplementary Figure 7b are the numerically calculated phase masks with the illuminating beam coming from the source. These numerically calculated masks are calibrated and transformed onto the surface of the SLM to become Supplementary Figure 7a. For the numerically calculated masks, the $x$-axis represents frequency, and the $y$-axis represents Fourier space ($k_y$) with respect to the 1D input array of the MPLC. For the results shown in this work, the masks are calculated in the spectral domain, and the resulting field are shown in Supplementary Figure 7c. Each spectral component of Supplementary Figure 7b has been Fourier transformed to reveal the distribution of amplitude and phase in the spectral domain at the input to the MPLC. This plane is divided into two regions, a region of interest, corresponding with the positions of the 45 inputs of the MPLC device, and a "don't care" region, where light must be discarded in order to achieve the desired beam quality as mentioned above. The plane of Supplementary Figure 7c does not physically exist in the actual device, it is simply the plane in which the target field is computed. However for an alternate implementation of this device based on two planes of phase manipulation using two SLMs, this plane would correspond with a second SLM. Supplementary Figure 7d is the 2D Fourier transform of Supplementary Figure 6b and represents a physical plane in the device at the grating or at the input to the MPLC.

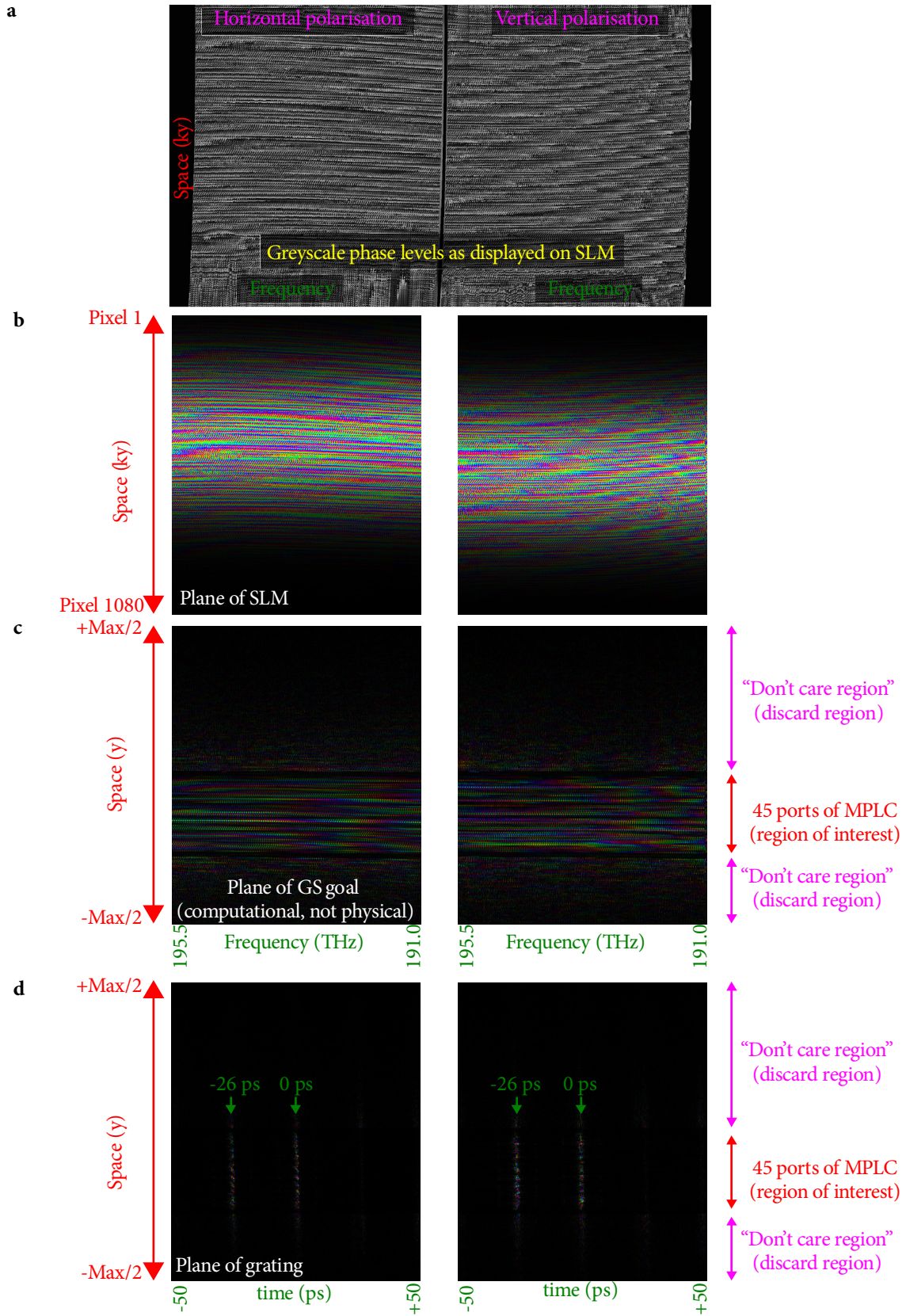

**Supplementary Figure 7 | Example phase masks and corresponding fields in the spectral and temporal domains for the generation of two delayed spots at the end of the multimode optical fibre.**

# Supplementary Note 2 : Characterisation of spatially-dependent loss (47m35s)

From the measured optical transfer function of the device it is possible to extract the spatial and frequency-dependence of properties such as loss. Another advantage of the full linear description obtained through a measurement of the frequency-dependent transfer matrix of the device is that all possible linear properties can be extracted. Properties such as the loss of each individual mode/port as addressed in the MPLC array, but also absolute worst-case performance metrics can be extracted such as mode-dependent loss (MDL) and averages over all possible combinations of modes such as insertion loss (IL). MDL, also known as the condition number of a matrix, is a measure of how "invertible" the matrix is. It is the ratio between the maximum and minimum singular value of the matrix and represents the maximum possible difference in loss between any two superpositions of states through the device. A detailed discussion and explanation of the singular value decomposition and the concepts of insertion loss and mode dependent loss as they relate to optical devices such as these have been previously discussed[6,13] (https://www.youtube.com/watch?v=TqAr4UFtoqw&t=45m50s). The insertion loss of the device can be seen in Supplementary Figure 8a. At the centre wavelength there is 17 dB of insertion loss. Which consists of approximately 3 dB of polarisation-dependent loss due to the static splitting of the horizontal and vertical components of the beam onto separate sides of the SLM. As loss is defined here, when a given port of the MPLC is being selected by the spectral pulse shaper in a given polarisation, the power in the orthogonal polarisation is being considered loss. This is a worst-case assumption, as in practice most input states contain components of both horizontal and vertical components, or the input polarisation could be aligned to the dominant axis to minimise loss. An alternate design for this device would address all 90 spatial/polarisation modes as a single 1D array as seen by the SLM. That is, no polarisation splitting onto the SLM surface, only polarisation recombining near the MPLC. The advantage of such an approach would be a reduction in worst-case loss, at the expense of higher requirements on the spatial resolution of the SLM and its associated phase masks. The MPLC device has approximately 5 dB of loss. This consists of approximately 2 dB due to the reflectivity of gold over 14 reflections, 1 dB of theoretical loss, and a remaining 2 dB of coupling loss and other imperfections. The 45-port spectral pulse shaper incurs the remainder of the loss at approximately 9 dB. For context, commercially available multi-port spectral pulse shapers such as a Finisar Waveshaper 16000A (19-port) are currently specified as 5 dB of insertion loss.

The measured mode dependent loss for the device is relatively good at between 7.2 dB for 193.819 THz and 10.4 dB at 195.409 THz. Whilst minimising insertion loss is important for applications that are power-limited, MDL is the more important measure with respect to the beam quality and the number of spatial channels the device can truly address. A high MDL means the transfer matrix of the system is difficult or impossible to invert and some spatiotemporal states may not be accessible, as the required spatial states are effectively missing. As opposed to MDL, which is calculated using the singular value decomposition and corresponds with superpositions of the modes the transfer matrix was measured in, Supplementary Figure 8b illustrates the losses in terms of the ports of the MPLC addressed by the spectral pulse shaper, a total of 90 ports/modes corresponding with 45 spatial modes in both polarisations. The differential port loss for a given frequency is fairly consistent at approximately 5 dB, with a roll-off of approximately 4 dB across the frequency band, as can also be seen in the insertion loss plot of Supplementary Figure 8a. Taken together, the lowest loss port at the lowest loss frequency has approximately 9 dB better loss than the worst port at the worst frequency. Similar information is presented in Supplementary Figure 8c as the total loss of each port at each frequency. The 45 modes located at the top of Supplementary Figure 8c correspond with the 45 ports in the horizontal polarisation, and the bottom 45 modes correspond with the same 45 ports in the vertical polarisation. For example, mode 1 and mode 46 are addressing the same port of the MPLC, from the left and right sides of the SLM respectively, corresponding with the two orthogonal polarisation components. The ports are enumerated in order of the Hermite-Gaussian modes they address at the output of the MPLC, rather than their 1D positions at the input to the MPLC. That is, ports 1 and 2 are not physically adjacent spots at the input to the MPLC, they are adjacent modes in Hermite-Gaussian space. Mode 37 and the corresponding mode in the orthogonal polarisation (mode 82) are the worst ports of the MPLC device, particularly at the lower frequencies, which can also be seen as the light blue and red traces of Supplementary Figure 8b.

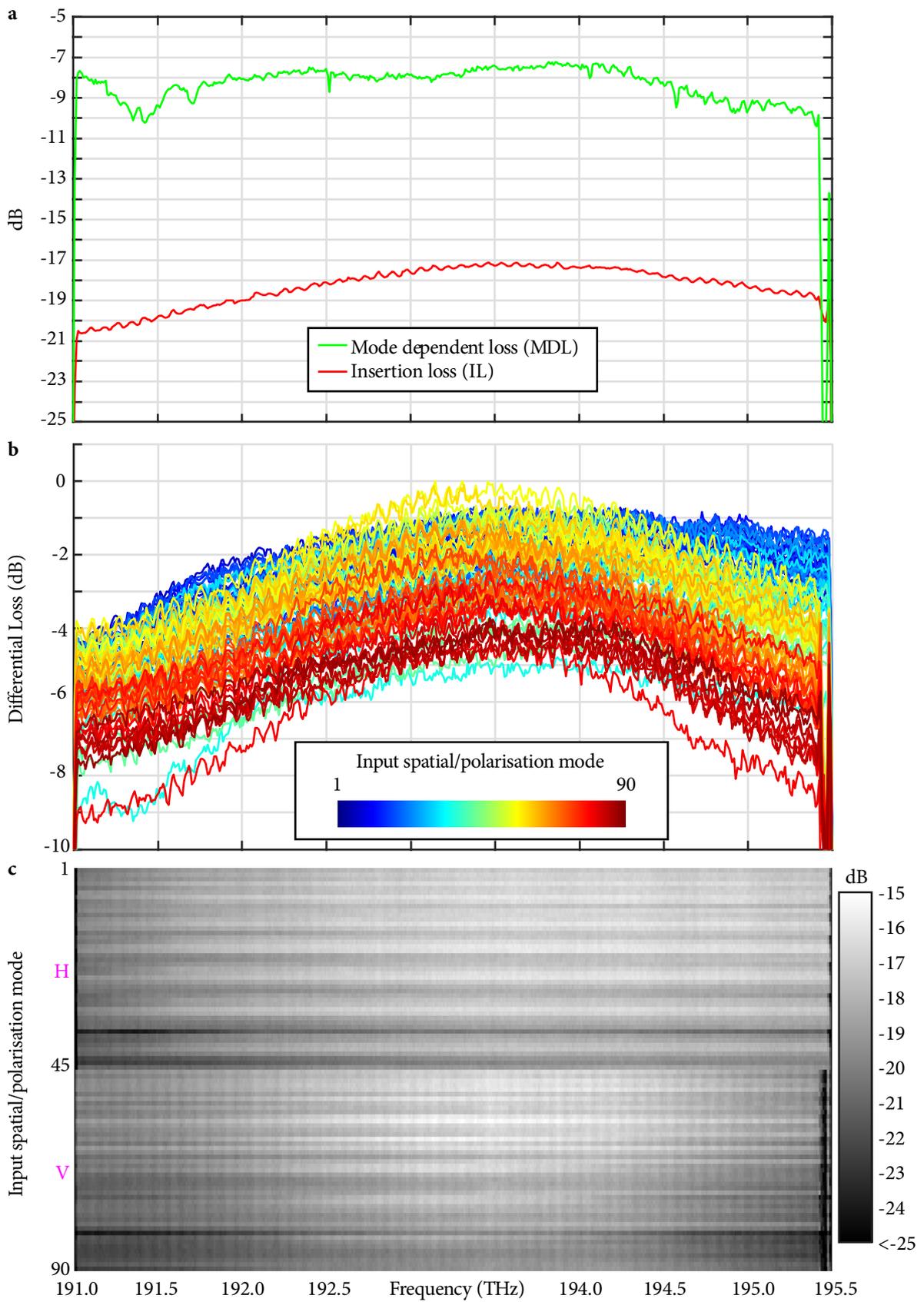

**Supplementary Figure 8 | Characterisations of the spatial and frequency dependent losses of the device. a,**

Insertion loss and mode-dependent loss. **b,** Differential loss between HG modes excited at the input of the fibre as seen from the output of the fibre. **c,** Total loss between HG modes excited at the input of the fibre as seen from the output of the fibre.

## Supplementary Note 3 : Characterisation of impulse response (55m40s)

Supplementary Figure 9 illustrates aspects of the temporal performance of the device, including the attached MMF. Supplementary Figure 9a is the impulse response obtained from the measured transfer matrices. For each spatial HG mode in each polarisation as addressed at the input of the MMF, the corresponding impulse response is shown. The total impulse response averaged over all modes in both polarisations is illustrated in Supplementary Figure 9b. This is the spatiotemporal response of the system as a whole and includes contributions both from the MMF, as well as the spectral pulse shaper sub-system. The differential group delay and chromatic dispersion within the spectral pulse shaper is primarily between the two orthogonal polarisation components. The MPLC sub-system should have negligible modal dispersion. In total there is approximately 2 ps of differential delay with approximately equal amounts of delay introduced within the spectral pulse shaper itself, as well as the MMF.

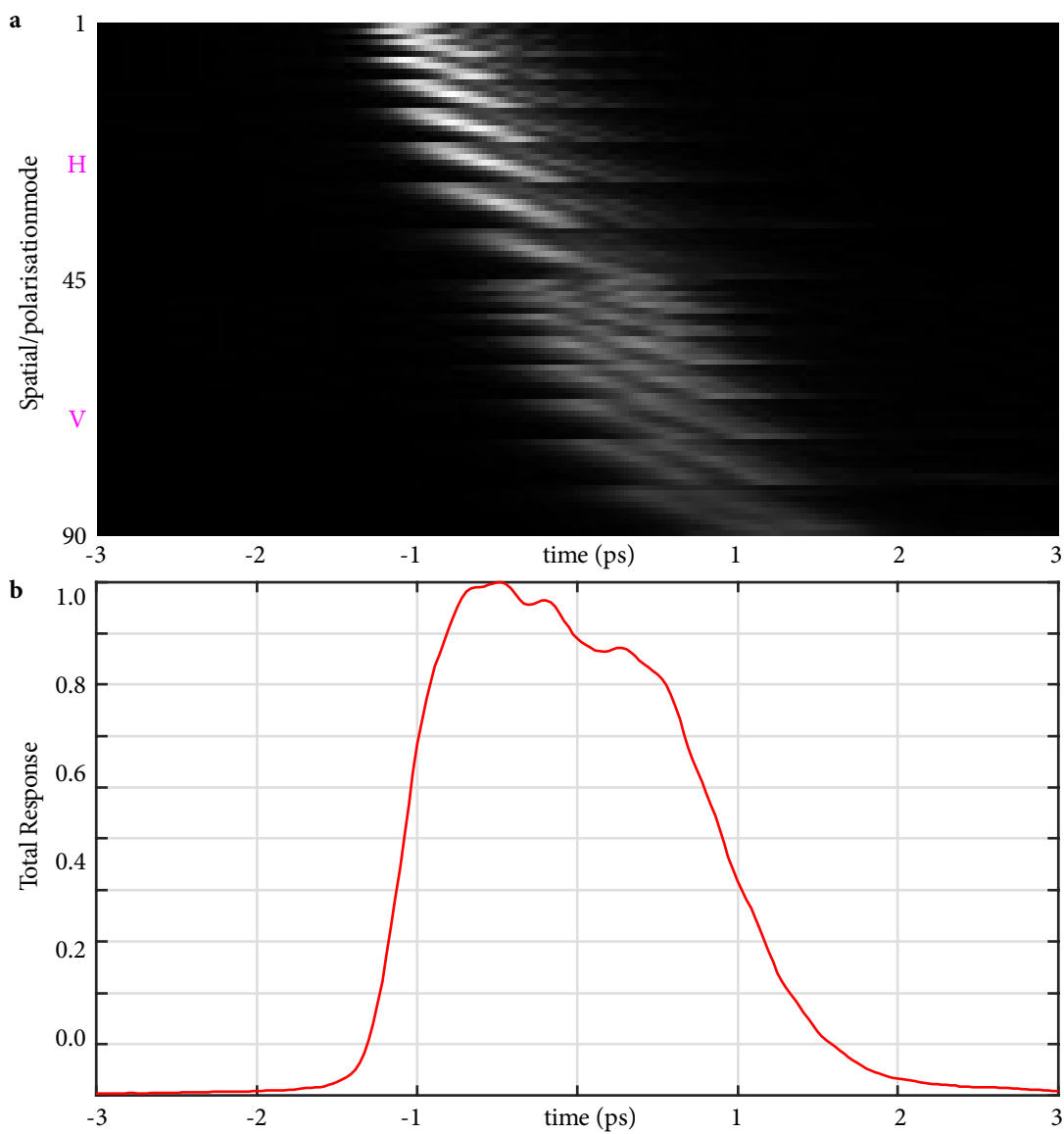

**Supplementary Figure 9 | Characterisation of the impulse response and delay-dependent loss of the device. a,** Measured impulse response for each input HG mode in each polarisation. **b,** Total impulse response over all spatial/polarisation modes, this is a characterisation of the uncompensated optical system.

## Supplementary Note 4 : Characterisation of delay-dependent loss (56m52s)

Supplementary Figure 10 illustrates a characterisation of the delay-dependent loss of the device. There are multiple mechanisms by which delay-dependent loss are introduced within such a device. First, there is the delay-dependent tilt of the beam, which reduces coupling into the MPLC. A delayed beam path will be spatially offset in the plane of the grating, which results in a corresponding angle of incidence when Fourier transformed into the plane of the MPLC input. Second, there is a delay-dependent defocus of the beam associated with the fact that it has physically travelled a longer path. Third, and the most significant contribution here, is the diffraction efficiency of the SLM. Creating light at large delays involves the steering of light to large angles across the spectral axis of the SLM. The diffraction efficiency of an SLM decreases as the diffraction angle is increased. This is partially due to quantisation and pixelisation effects, but largely due to the non-ideal performance of liquid crystals[14]. In the test of Supplementary Figure 10, a horizontally polarised focused spot is created at the output of the MMF at a delay of 0 ps. In the vertical polarisation at the MMF output another focused spot is created at a different spatial location. The delay of this vertically polarised spot is stepped from -26 ps to 26 ps in 2 ps steps. The measured fields in both polarisation components at the output of the fibre for each trial can be seen in Supplementary Figure 10. The *x*-axis of these images represent the spatial field at a particular delay for a given phase mask on the SLM, in the same manner as examples such as Fig. 3 of the main document. The *y*-axis represents different phase masks designed to generate a vertically polarised spot at a desired delay. The results of Supplementary Figure 10a are normalised with respect to the case where the spots in both polarisations arrive at the same delay of 0 ps. Hence for all other delays, the ideal case would be a response of 0.5, as naturally the intensity in a given time interval is halved when the spots are temporally separated. It can be seen that there is asymmetry in the delay-dependent loss, with negative delays suffering less loss than positive delays of the same magnitude. The exact mechanisms that contribute to this asymmetry cannot be definitely measured, however one aspect could be the differing diffraction efficiency of the SLM when steering in one direction versus the other[14]. Other mechanisms could be related to issues of optical alignment such as defocus into the MPLC input or clipping of the beam.

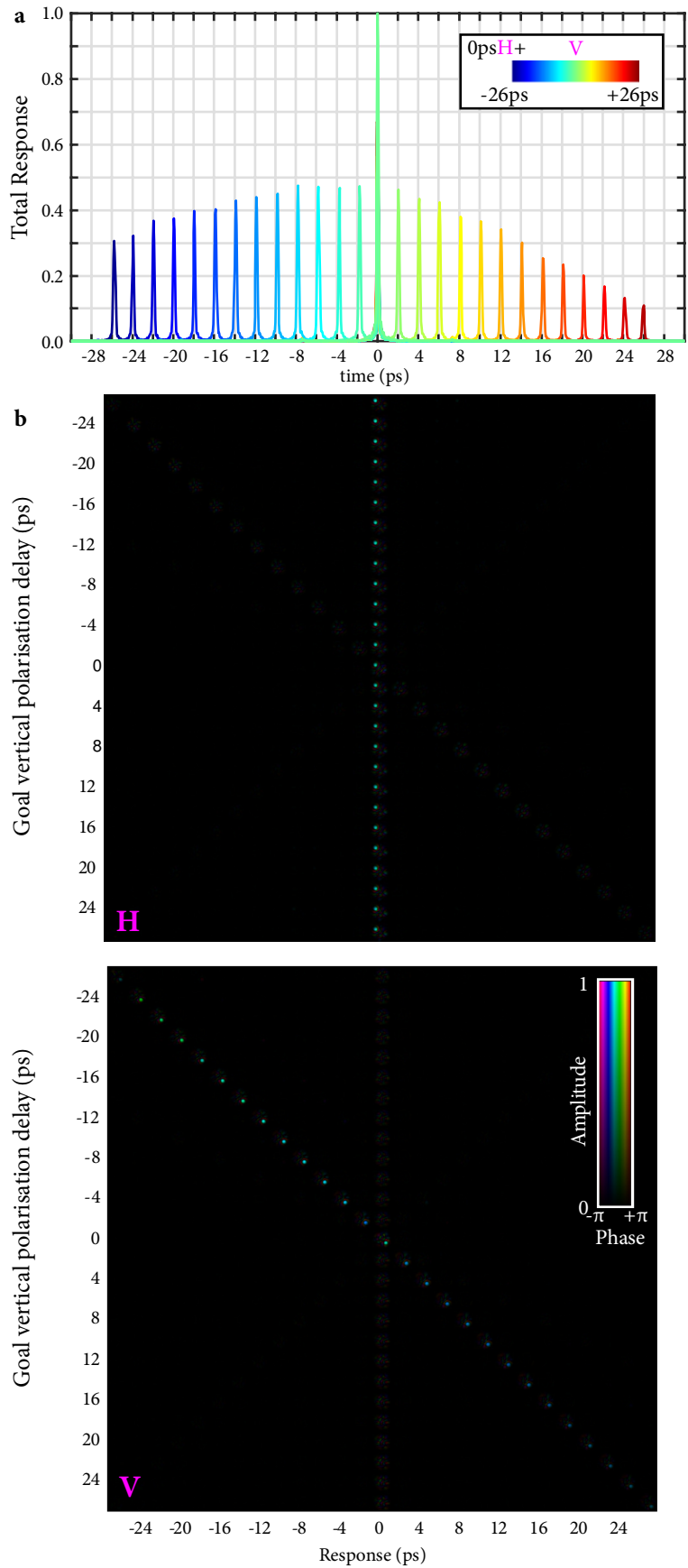

**Supplementary Figure 10 | Measured fields for characterisation of delay-dependent loss.**

# Supplementary Note 5 : Spatiospectral stress test (59m26s)

The tests of Supplementary Figure 11 to Supplementary Figure 15 are specifically designed to be challenging for the device to implement. Tests which require all spatial and polarisation modes to be excited with equal power by the SLM across the entire operating frequency bandwidth of the device, with rapid changes in mode superposition from one frequency component to the next. That is, tests which require the SLM to address dense combinations of the spatial and temporal degrees of freedom simultaneously with a single phase mask. The spatiotemporal states these represent are not likely to be of practical use, or likely to occur in any realistic setting. They are contrived as a means of testing the limits of the device in worst-case scenarios. The tests consist of implementing columns of the discrete Fourier transform (DFT) matrix in spectral bands of between 10.0675 GHz and 50.3375 GHz. Corresponding with approximately 2 to 10 pixel widths across the spectral axis of the SLM. The significance of the DFT matrices here is that they require the SLM to generate all 90 spatial and polarisation modes with equal amplitude across the entire operating frequency bandwidth of the device (4.4 THz), with each frequency band (10.0675 to 50.3375 GHz) being orthogonal to all adjacent and nearby bands. This is a particularly challenging scenario which requires phase masks that have high spatial frequencies in both the spatial and spectral directions across the entire clear aperture of the device. This state is being generated by the SLM in some plane roughly corresponding with the input of the MMF. However it is the corresponding output state that is experimentally accessible and will be measured and tested against. Attempting to generate the same DFT spatiospectral state at the output of the MMF would in fact be easier, as the scattering in the fibre would make it very unlikely that all HG modes would need to be equally excited at all frequencies. The DFT matrix is implemented in the basis of the Hermite-Gaussian modes (output of the MPLC). Implementing in the basis of the ports of the 1D array facing the spectral pulse shaper (input of the MPLC) would be even more difficult, as the mismatch between the Fourier transform of the 1D array and the source beam coming from the laser on the SLM would be always be at its worst, across all frequency components, requiring a large amount of loss. This worst-case scenario also occurs in the HG basis for frequency bands where all modes are required to be generated in-phase. A scenario where the DFT is being implemented in both the 1D basis of the spots and the 2D basis of the HG modes simultaneously.

Supplementary Figure 11a illustrates the theoretical spatiospectral state being generated by the SLM. For the case of 50 GHz bands, there are coincidently as many spatial/polarisation modes as there are frequency bands across the 4.5 THz of the device. For all other bandwidths, illustrated in Supplementary Figure 12 to 15, the DFT matrix repeats across the frequency axis. Supplementary Figure 11b are the theoretical coefficients for the corresponding phase mask calculated as per the description above. Each spectral band is calculated to have a beam quality of 95%, but no attempt is made to equalise the overall loss of each frequency band. The differences in loss are inherent for a single plane hologram as discussed above. The worst-case can be seen at the far right of Supplementary Figure 11b, where all modes must be generated in phase. This scenario corresponds with the Fourier transform of the 1D spot array focusing to a single small spot on the SLM. This small spot is much smaller than the source beam coming from the laser. Theoretically due to this size mismatch, the maximum possible coupling between the two would be 14%. Supplementary Figure 11c is the corresponding output spatiospectral state that would be expected if we propagated the ideal state of Supplementary Figure 11a through the measured transfer matrices of the device and MMF. Supplementary Figure 11d is the measured output spatiospectral state. The complicated nature of the output states makes comparison by eye of Supplementary Figure 11c and 11d difficult. However it is summarised in Supplementary Figure 11e, as the normalised field correlation squared between the expected and measured spatial/polarisation output states as a function of frequency. That is, the proportion of the power that is measured to be in the expected spatial state at each wavelength. Ideally this would be 1.0 for all frequencies, or more realistically, 0.95 for all frequencies as that was the specified quality for convergence in the Gerchberg-Saxton algorithm used to calculate the phase masks. The green series of Supplementary Figure 11e follows the centre frequencies of each band and is relatively flat at approximately 0.9 from 191.0 THz to 195.4 THz. For this particular test, the worst-case in-phase spectral band is just outside the operating bandwidth of the device. The red series is the full measurement at the resolution of the swept-wavelength digital holography system (10.0675 GHz). The dips in response are to be expected, as these correspond with the spatial/polarisation state transitioning between two orthogonal states from one frequency band to the next. This transition must occur over some frequency bandwidth due to the physical size of the beam on the SLM along the spectral axis, as well as the finite spatial resolution of the SLM itself. Supplementary Figure 11f is the total impulse response of the ideal and measured output states corresponding with Supplementary Figure 11d and 11e. There is good correlation between the two, especially at the smaller magnitude delays, consistent with the delay-dependent loss characterisation of Supplementary Figure 9c.

In Supplementary Figure 12 to 15 it can be seen that the correlation between the ideal and measured states in either the spectral or temporal domains, are well correlated for the 40 and 30 GHz bands at mostly in excess of 80%. Some degradation starts to be observed at 20 GHz, and quality is poor for 10 GHz. This is to be expected as the size of the beam on the SLM is approximately 15 GHz in the spectral direction. Looking closely at the spectral correlations, it can also be seen that there are dips in correlation for the green series at frequencies which correspond with the worst-case of in-phase mode generation. Testing narrower bands in the frequency domain, corresponds with testing longer delays in the time domain. This can be seen in the temporal responses of Supplementary Figure 11f to 15f, as an increasingly long impulse response as the spectral bands become narrower. In the time domain plots of Supplementary Figure 11f to 15f, there is a high correlation between the ideal and measured responses within the delay interval of approximately -20ps to +10ps. For delays outside this range, performance degrades as the finite spectral resolution in the plane of the SLM, makes these larger delays difficult to obtain accurately. This is consistent with the delay-dependent loss characterisation of Supplementary Figure 9c.

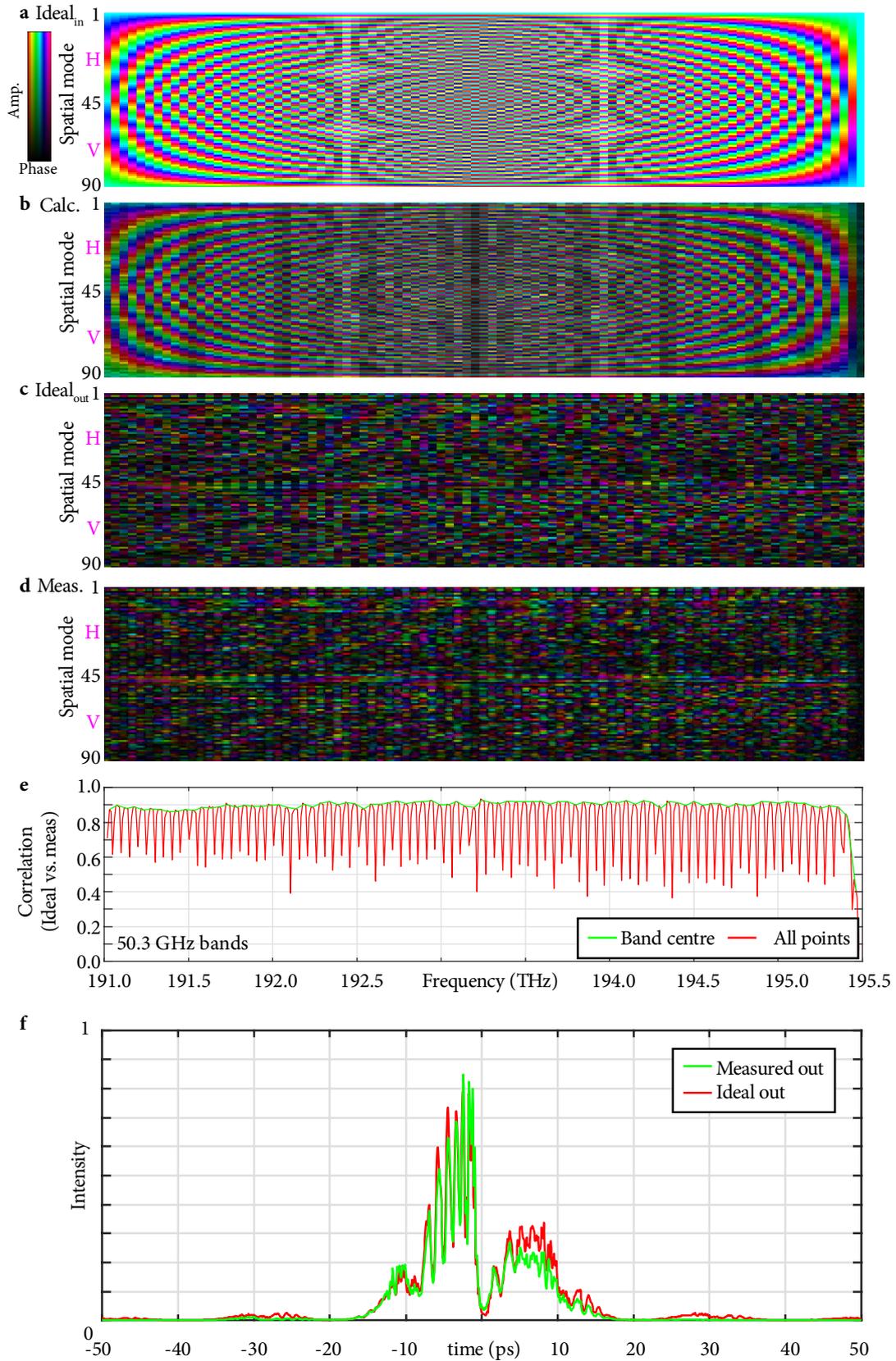

**Supplementary Figure 11 | Characterisation of discrete Fourier transform spatiospectral states generated by the SLM for 50 GHz spectral bands.**

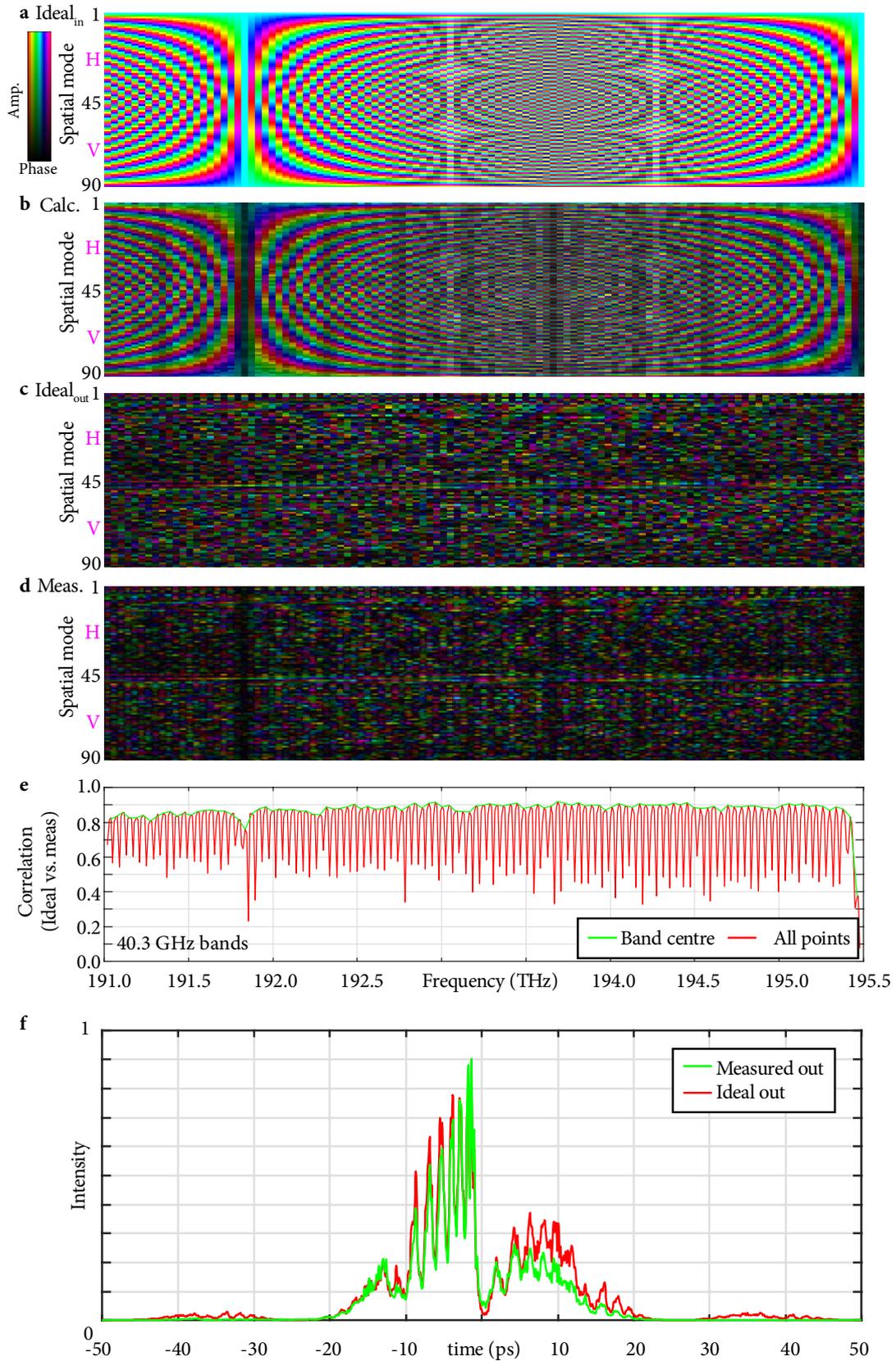

**Supplementary Figure 12 | Characterisation of discrete Fourier transform spatiospectral states generated by the SLM for 40 GHz spectral bands.**

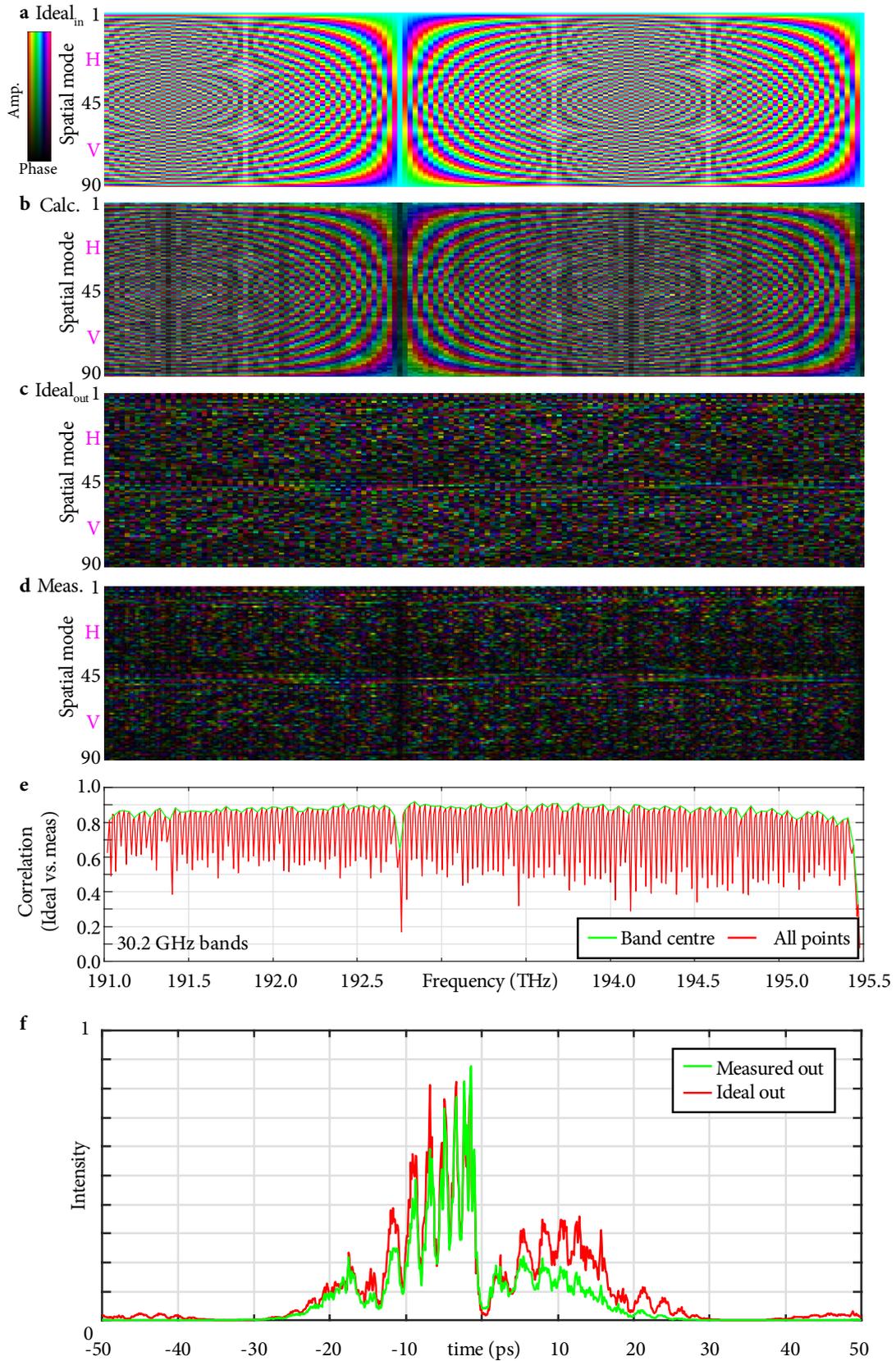

**Supplementary Figure 13 | Characterisation of discrete Fourier transform spatiospectral states generated by the SLM for 30 GHz spectral bands.**

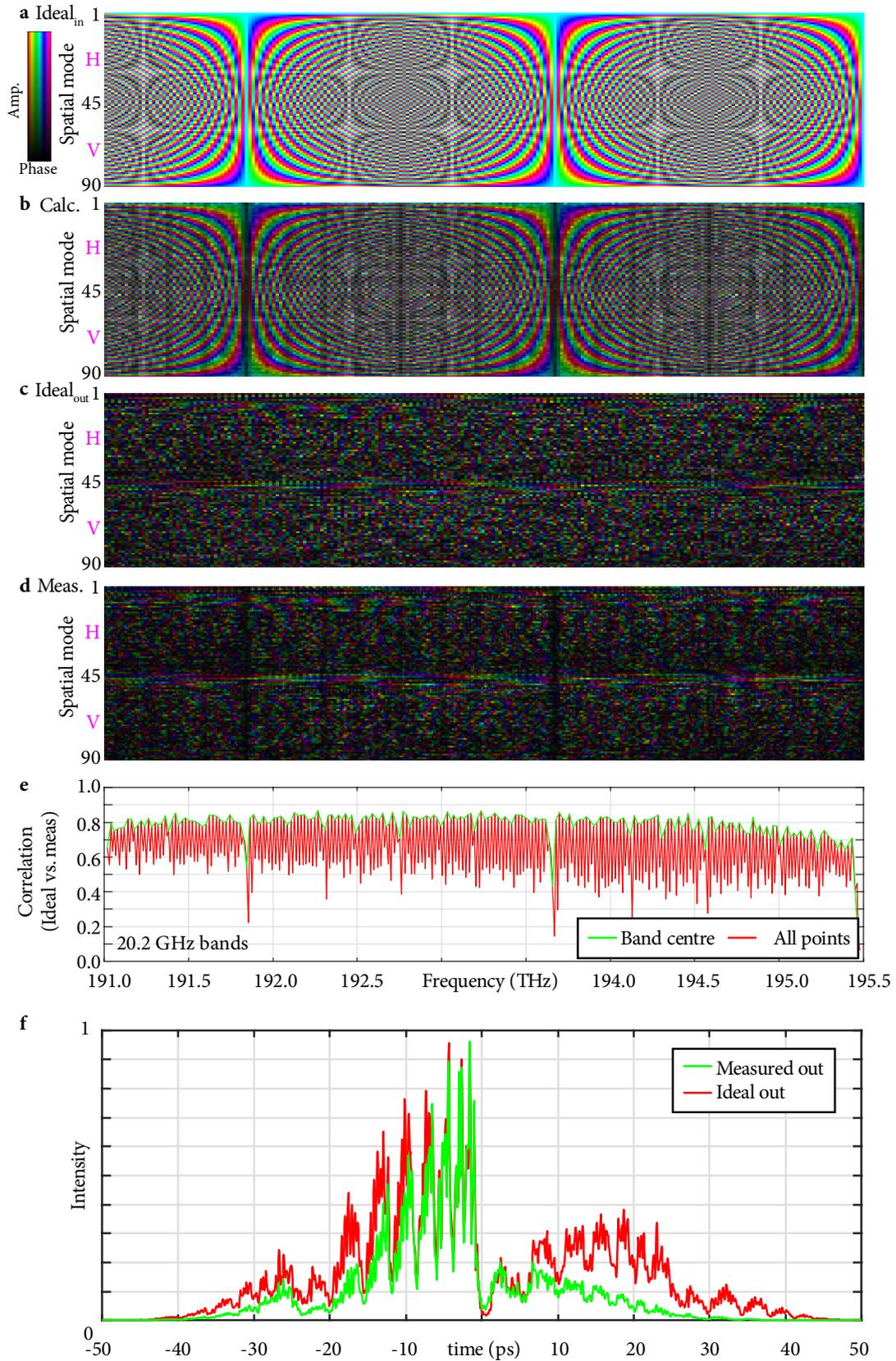

**Supplementary Figure 14 | Characterisation of discrete Fourier transform spatiospectral states generated by the SLM for 20 GHz spectral bands.**

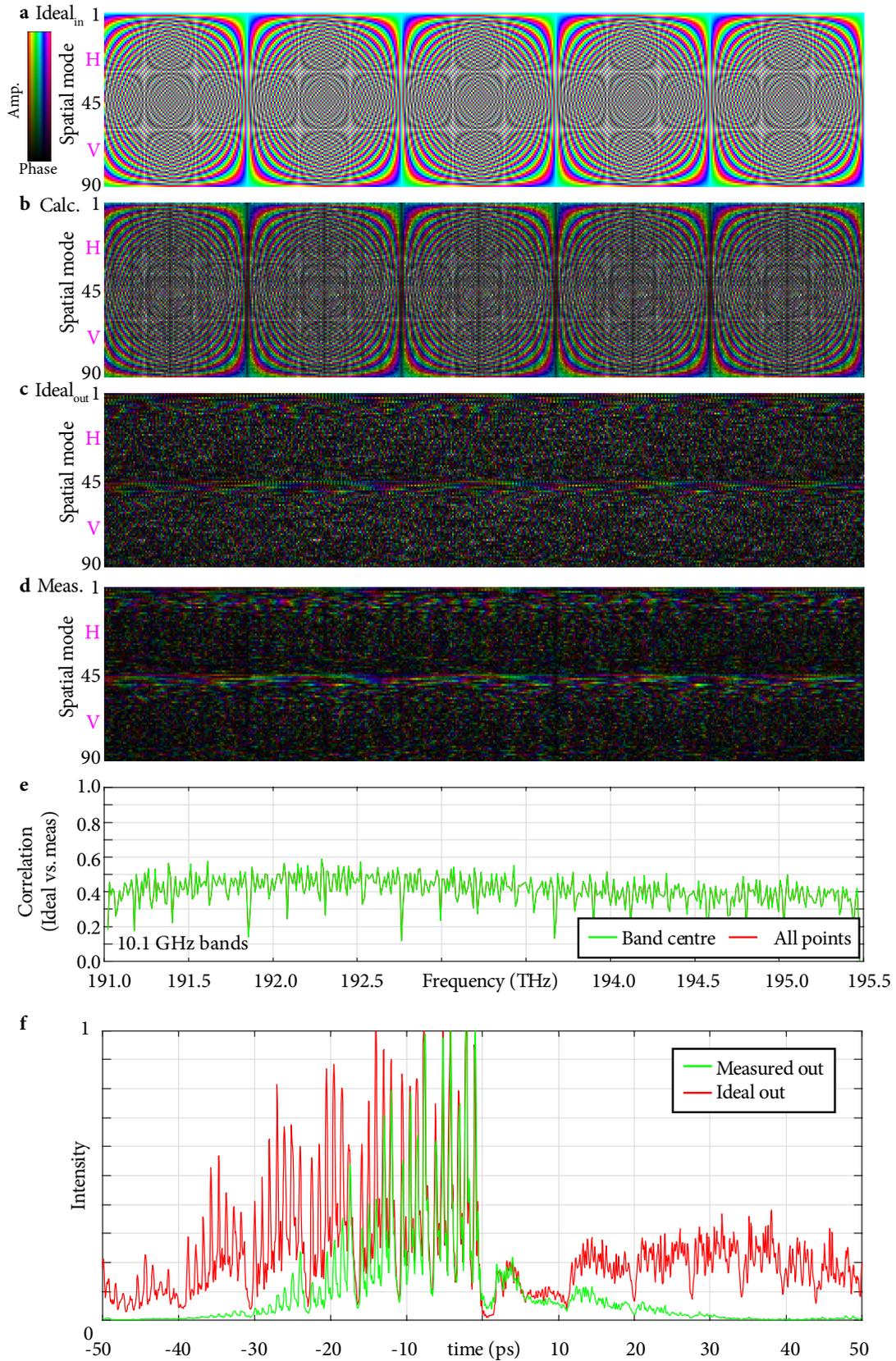

**Supplementary Figure 15 | Characterisation of discrete Fourier transform spatiospectral states generated by the SLM for 10 GHz spectral bands.**

# Supplementary Note 6 : Spatial and temporal intensity enhancement example

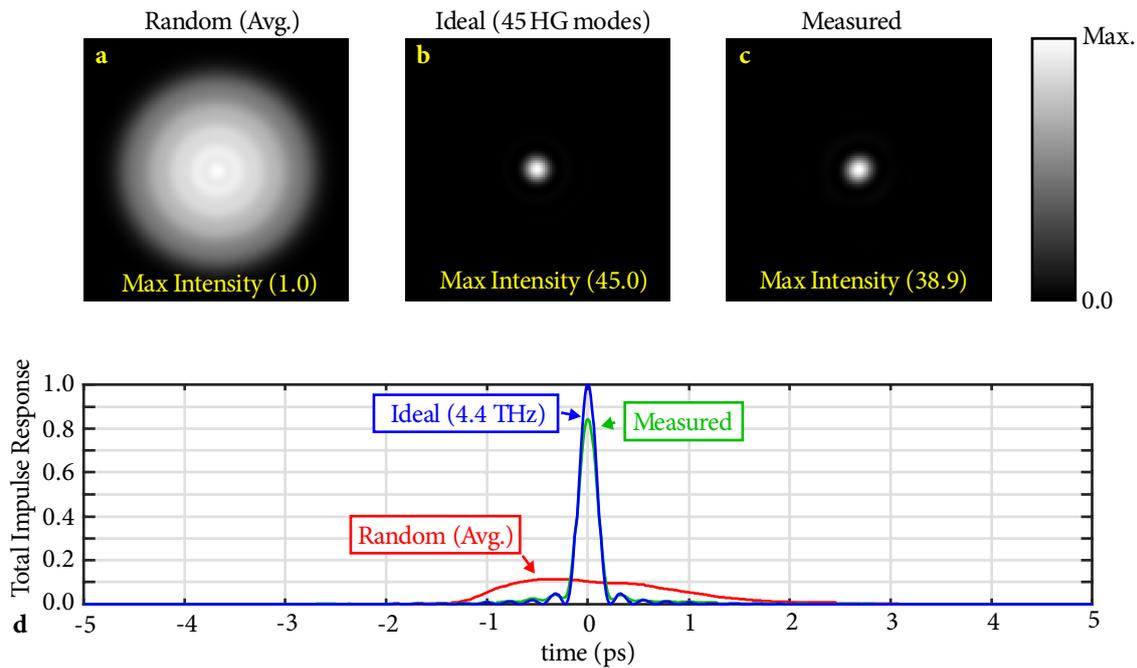

**Supplementary Figure 16 | Example of spatial and temporal intensity enhancement. a,** Average spatial intensity over all random output spatial states. **b,** Output spatial state with theoretical peak intensity possible using 45 HG modes. **c,** Measured output spatial state generated at end of MMF. **d,** Impulse response for average random output spatial state (red), theoretical peak intensity for bandwidth limited pulse (blue) and measured impulse response (green).

In this demonstration, a 45 degree linearly polarised, diffraction-limited, and bandwidth-limited spatiotemporal focus is generated at the end of the MMF. The corresponding intensity enhancement is then compared with a random output state, as well as the theoretical maximum possible intensity given the number of spatial modes and spectral bandwidth supported by the system. Supplementary Figure 16a is the reference average random output spatial state. Averaged over all possible superpositions of HG modes, which is equal to the sum of the intensities of each of the 45 HG basis spatial modes. Supplementary Figure 16 is a diffraction-limited spot as approximated by a superposition of the 45 HG basis modes. This is theoretically the most intense focus which can be created using the basis modes. Supplementary Figure 16c is the corresponding spatial state measured at 0 ps at the output of the MMF. When the spatial states are normalised to unity total power, the peak intensity of the spot is 38.9 times the average random state, approximately 86% of the theoretical maximum of 45. In this comparison polarisation has been averaged, meaning it does not contribute to the measure of intensity enhancement. If the metric was instead defined as the intensity enhancement at the output of the MMF in a given polarisation, an additional factor of two would be gained as the average random state power would be divided amongst two polarisations.

Supplementary Figure 16d is a similar illustration to Supplementary Figure 16a-c in the temporal rather than spatial domain. The average random impulse response, illustrated in red, is over all spatial and polarisation modes with equal power at the output of the MMF. This is the temporal response of the spatial state of Supplementary Figure 16a. An impulse response similar to Supplementary Figure 9b, but normalised for equal output power rather than input power. The ideal response shown in blue is a sinc pulse, corresponding with the full bandwidth of the system (4.46 THz), and has a peak intensity at 0 ps that is 9.85 times higher than the average random state. The measured response shown in green has a peak intensity of 8.31 times the average state or 84% of the theoretical maximum for a bandwidth limited pulse.

## Supplementary Note 7 : Spatial/polarisation state fidelity

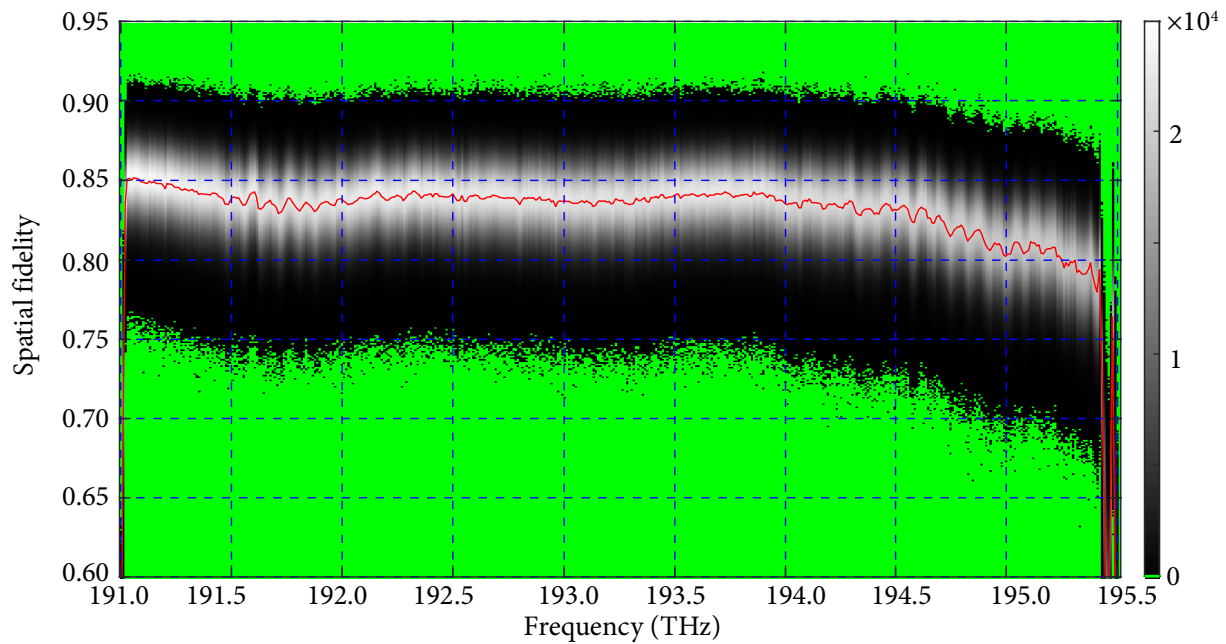

**Supplementary Figure 17 | Distribution of fidelity for random spatial/polarisation target states.** Given the measured transfer matrices for the device, one million random spatial/polarisation states per ~10 GHz spectral band are propagated back and forth through the system to yield a measure the achievable fidelity of spatial/polarisation states when maximising the power delivered to the target state, given the presence of mode dependent losses. Green indicates exactly zero states over all trials in that specific 0.001 wide histogram bin. Red line indicates average fidelity.

For the results presented in this work, a given target spatial/polarisation state is specified for each frequency component at the distal end of the multimode fibre. These states are numerically back-propagated using the conjugate transpose of the measured transfer matrices, to find the corresponding input state. For an ideal system, these transfer matrices would be unitary to within some overall loss common to all spatial and polarisation modes. All spatial/polarisation modes would have the same loss, and the conjugate transpose of the transfer matrix would also be its inverse. For such a system, ignoring any additional imperfections associated with generating superposition states as discussed previously in the section 'Spatiospectral stress test', a desired spatial/polarisation target state could be generated with maximum power and perfect fidelity. Theoretically, for an imperfect system with mode dependent loss (Supplementary Figure 8), "back-propagation" using the inverse of the transfer matrix will yield an input state capable of generating the output state with perfect fidelity, at the expense of some additional loss. Using the conjugate transpose of the transfer matrix for back-propagation as we do in this work, has the opposite trade-off. It yields the input state which would deliver the maximum amount of power to the target state, at the expense of overall fidelity.

Supplementary Figure 17 is a histogram which illustrates the degradation of spatial fidelity due to mode dependent loss when using numerical back-propagation by the conjugate transpose of the measured transfer matrices. For each ~10GHz band, one million random spatial/polarisation states are generated numerically. For a random state, each of the 90 modes is assigned a random amplitude between 0 and 1, and a random phase, and then normalised to unity total power. These states are then back-propagated through the multimode fibre to the input using the conjugate transpose of the transfer matrices, and then forward propagated using the transfer matrices to return to the distal end of the multimode fibre. This state is then renormalised to unity total power, and overlapped with the target state to yield the spatial fidelity. That is, the proportion of the output power which is in the target state. The results of Supplementary Figure 17 are linked to the results presented in Supplementary Figure 6 and 8, which also relate the mode dependent losses. The histogram of Supplementary Figure 17 indicates the number of random states per million trials per 10 GHz frequency band, which were found in each 0.001 wide fidelity bin. The mean fidelity within the 4.4 THz operating band is 0.83, and is indicated for each frequency with the red line. The distribution of fidelities is not

Gaussian, skewing towards low fidelities, but approximating as a normal distribution yields a standard deviation of 0.04. Highlighted in green, are histogram bins which recorded exactly zero states over all trials.

## Supplementary Note 8 : Frequently Asked Questions

**Isn't time reversal just phase conjugation?**

Sometimes the terms are used interchangeably, however using the terminology we prefer[15], time reversal represents broadband control, whereas traditional phase conjugation is CW or narrowband. Traditional phase conjugation has no temporal features and is spatial-only. Consisting of a single spatial wavefront for all delays. Time reversal described in the frequency domain is broadband phase conjugation[15,16]. Each spectral component has its own wavefront, which in turn creates spatiotemporal features. Time reversal is spatiotemporal, whereas CW phase conjugation is only spatial.

**How is this device different from previous wavefront shaping experiments?**

This is the first optical wavefront shaping device which can arbitrarily and independently control all light's degrees of freedom; 2D space, time/frequency and polarisation. There is no coupling between these degrees of freedom and control of each maps to a unique position on the SLM. For example, assigning a certain amplitude and phase to any spatial/polarisation mode, does not define the properties of any other spatial or spectral component.

There have been many demonstrations over the years with some partial ability to control the spatial, polarisation and temporal aspects of an output beam. However due to the way those devices work, some types of beams are not possible, as components of the beam that would need to be independently controlled are not independently controllable. Imagine tracing the different spatial/polarisation and spectral components of the target beam back through the optical system from the target to the wavefront controlling device(s), typically an SLM. For previous demonstrations, for an arbitrary target, there would be at least two components of the beam, for example spatial modes and spectral modes, that must be controlled independently in the plane of the SLM in order to generate the target, yet both end up mapped to the same position on the SLM. Attempting to control one, also controls the other, and the target field is not accessible.

For example, some demonstrations make use of a highly scattering media which strongly couples the spatial and temporal properties[17,18]. Hence, by controlling only the spatial properties at the input, both spatial and temporal properties at the output can be manipulated. However again, this is only partial control. For an arbitrary spatiotemporal output field, the corresponding time reversed input field will be spatiotemporal; a non-separable function of both space and time. Yet only the subset of target fields that are spatiotemporal at the output, but purely spatial at the input can be accessed by these methods. Our system does not rely on spatiotemporal coupling in the complex media itself to address the temporal degree of freedom. Hence the device can operate in uncoupled scenario like free-space, all the way to a completely coupled scattering media. The limitations of our system are in terms of the maximum amount of spatiotemporal detail (addressable number of spatial and temporal/spectral modes), but not in terms of the types of coupling scenarios (none, weak, strong) it can handle.

Another common trade-off is a system which uses some dispersive element like a grating, to map a spatial dimension to the spectral degree of freedom which is in turn controlled by the 2D surface of an SLM[19–21]. In these cases, one spatial degree of freedom is not independently controllable from the spectral degree of freedom. To take a specific example, if we look at Fig. 1d of Sun et al.[20], we can see that one spatial dimension at the sample plane, maps directly to one spectral dimension. Each spectral component incident on the sample has its own incident angle, and the two components, spatial and spectral, cannot be controlled completely independently. Thinking of the system in reverse, trace an arbitrary wavelength back from the sample at an arbitrary angle. Trace the same wavelength forward from the laser source. The forward and backward paths will not overlap on the SLM for an arbitrary incident angle at the sample. Our system is similar to these approaches, however it use the 1D-to-2D spatial transformation of the MPLC[3] to get around this limitation. The MPLC transformation allows the SLM to address the spatial degree of freedom as if it is a one-dimensional degree of freedom. The MPLC acts as a kind of 1D 'lookup table' from which 2D output beams can be selected from a 1D list. Our device generates beams which are spatiotemporal in both the near and far-field. Similar to McCabe et al.[19] where the spectral components are recombined with a grating, but different from Sun et al[20] and Hernandez et al.[21] in which the spectral components are recombined using a lens and hence are only temporally short at the focus.

**Why not physically back-propagate your target field?**

Our approach is similar to techniques such as DORT[22,23] in time reversed acoustics, or transfer matrix based phase conjugation approaches for CW light[24]. Numerical back-propagation using a transfer matrix has many advantages over physical back-propagation, especially for sophisticated beam types. A numerically back-propagated beam is more experimentally accurate as it eliminates the source of error associated with physically generating complicated fields. For relatively simple beams types, which have the same spatial profile at all delays (space-time separable), such as spatiotemporal focus, physical generation is relatively straightforward and could be done using a single-mode fibre, perhaps coupled to a traditional spectral pulse shaper. However part of the point of this paper is the generation of arbitrary vector spatiotemporal beams. Not only the relatively simple separable beam types, but rather beams which can have any value of amplitude and phase in any polarisation state, at any position at any point in time. These are the class of beams required to create time reversed waves in the general case, and a class of beams not previously demonstrated. Generating these beam types is considerably more difficult than space-time separable beams like spatiotemporal focus, and our system presented here is the first system capable of doing so. Hence to physically back-propagate these beam types, two copies of our optical systems would need to be built; one to physically generate back-propagated waves, and one to generate forward-propagated waves. Even so, if there was an advantage to taking this approach, building two systems could be justified. However as the numerical back-propagation approach using a transfer matrix is more experimentally accurate, allows any field to be delivered to the target from a relatively small number of measurements, and is also applicable to non-reciprocal system such as fibre amplifiers containing isolators, there is not a strong use-case for physical back-propagation. Numerical back-propagation is more accurate, has less optical components and is applicable in more experimental scenarios.

**Why do you generate your spatiotemporal fields in the frequency domain rather than time domain?**

Ultrafast pulse manipulation in optics is typically performed in the frequency domain, as the bandwidths required are beyond the capabilities of electronics. Many applications for which we'd expect a device such as this to be useful, would have an optical input from a source such as a pulsed laser.

An alternative approach would be a time domain implementation, perhaps using large numbers of Mach Zehnder modulators driven electrically. This would have similarities to a mode division multiplexing type system in optical telecommunications[8], but where the phase between spatial modes is carefully controlled at the transmitter and processing is performed optically at the transmitter side, rather than digitally at the receiver side. An advantage of this approach would be the ability to generate very long impulse responses, limited mostly by computer memory. The disadvantages would be comparatively low bandwidth (GHz rather than THz), and a potentially very complicated optical system. A time domain implementation in optics would be similar to a mode division multiplexing transmission system in optical fibre communications[8], but with special attention to the phase relationship between spatial channels.

**Why do you characterise in the frequency domain not the time domain?**

As mentioned above, our system implements its functionality using a frequency domain approach. As such, it makes sense to characterise the device in that same domain. Our optical characterisation apparatus based on swept-wavelength digital holography, is similar to a multi-port vector network analyser (VNA) in microwave systems, with all the usual advantages when it comes to impulse response measurements. Specifically, it has superior dynamic range, more accurate calibration and large flat bandwidth. In the frequency domain, we measure 444 CCD images over a span of 4.46 THz. One image for each spectral slice. For a spatiotemporal focus type demonstration, each of these spectral slices has the same power and uses the full dynamic range of the camera. The equivalent time domain characterisation would consist of a sequence of 443 blank images, with a single frame containing the information of interest. In terms of calibration, similar to a VNA, corrections are frequency dependent rather than delay dependent which increases experimental accuracy when operating in the frequency domain. For example, removing the dispersion difference between the reference arm of the interferometer and the device-under-test is trivial in the frequency domain. An additional advantage is that each measurement in the frequency domain also corresponds with a particular position on the SLM in the device itself. This is particularly important for building and debugging devices such as these.

Having said all that, characterising in the frequency domain is not always an option, and time domain can become necessary. This is either because the system being characterised is nonlinear, or because of other experimental considerations. For example, if operating at 800 nm, a high quality swept-frequency laser source may not be available, yet a high-quality pulsed source is. Leaving no choice but to characterise in the time domain.

**This system uses a multimode optical fibre, is it applicable to free-space/other scattering/complex media?**

Yes, in much the same way that a traditional single-mode fibre attached spectral pulse shaper can be used in either context. Designing a fibre-coupled device naturally results in free-space compatibility, however a free-space device is not automatically compatible with fibre. Stated another way, free-space supports all the spatial modes a fibre does, but a fibre does not support all the modes free-space does. A convenient aspect of our device is it addresses the spatial dimension of the beam in the Hermite-Gaussian basis. Which makes it well suited to both fibre and free-space applications. HG modes are not only eigenmodes of parabolic index fibres in the weak-guidance approximation, but also eigenfunctions of the Fourier transform, solutions to the paraxial wave equation and eigenmodes of laser resonators. As opposed to a spatial basis such as grids of spots in real or $k$-space, the HG basis allows light to be focused anywhere in the near or far-field, without dead-space. The HG basis also losslessly couples to fibres, whereas grids in real or $k$-space do not. It is also a more efficient basis for addressing a finite aperture[25] compared with simple grids.

Traditional spectral pulse shapers often have single-mode fibre outputs, which acts as a convenient conduit through which the light can be routed to the part of the optical system, where it is ultimately required. It is a fibre-coupled device, applicable to both free-space and fibre applications. Our MMF attached spectral pulse shaper can be thought of in a similar fashion. Just as some additional dispersion compensation may be applied to a traditional spectral pulse shaper to compensate for the attached SMF, additional spatiotemporal dispersion compensation can be applied to our spectral pulse shaper to compensate for the attached MMF if required. As with a traditional single-mode spectral pulse shaper, this assumes the power levels involved are compatible with fibre and are not sufficient to cause unwanted nonlinear effects. However the thresholds would be much higher in this case as due to the use of MMF rather than SMF.

The compatibility with optical fibre could also be important for future applications of spatiotemporal beam shaping for nonlinear optics.